\documentstyle[12pt]{JHEP3}
\newcommand{\bej}[1]{ \begin{equation}\label{#1} }
\newcommand{\eej}{\end{equation}}
\newcommand{\beaj}[1]{\begin{eqnarray}\label{#1} }
\newcommand{\eeaj}{\end{eqnarray}}
\newcommand{\eq}[1]{(\ref{#1})}
\def\ZZZ{{\hskip-3pt\hbox{ Z\kern-1.6mm Z}}}
\def\zzz{{\hskip-3pt\hbox{ z\kern-1mm z}}}

\newcommand{\be}{\begin{equation}}
\newcommand{\ee}{\end{equation}}
\newcommand{\ben}{\begin{eqnarray}\displaystyle}
\newcommand{\een}{\end{eqnarray}}

\newcommand{\sectiono}[1]{\section{#1}\setcounter{equation}{0}}

\def\one{{\hbox{ 1\kern-.8mm l}}}
\def\zero{{\hbox{ 0\kern-1.5mm 0}}}

\title{
Giant magnons in the D1-D5 system}
\author{Justin R. David\footnote{On lien from Harish-Chandra Research Institute, Allahabad.} $^{a}$ and Bindusar Sahoo$^b$\\
$^a$Centre for High Energy Physics,\\
$\;$Indian Institute of Science, \\
$\;$Bangalore 560012, India.\\
$\;$\email{justin@cts.iisc.ernet.in}\\
$^b$Harish-Chandra Research Institute, \\
$\;$Chhatnag Road., Jhunsi, \\
$\;$Allahabad 211019, India.\\
$\;$\email{bindusar@hri.res.in}
}

\abstract{
We study giant magnons in  the
the  D1-D5 system 
from both the boundary CFT and as  classical solutions of 
the string sigma model in $AdS_3\times S^3\times T^4$. 
Re-examining earlier studies of the symmetric product
conformal field theory 
we argue  that giant magnons in the symmetric product
are BPS states in a centrally extended
$SU(1|1)\times SU(1|1)$ superalgebra with two more 
additional central charges. The magnons carry these
additional central charges locally 
 but globally they vanish. 
 Using a spin chain
description of these magnons and the extended superalgebra we 
show that these magnons obey a  dispersion relation which is periodic in momentum. 
We then identify these states on the string theory side and show that 
here too they are BPS in  the same centrally extended algebra
and obey the same dispersion relation which is periodic in momentum. 
This dispersion relation arises as the BPS condition for the extended 
algebra and is 
 similar to that of magnons in ${\cal N}=4$
Yang-Mills}

\begin{document}
\baselineskip 3.5ex
\section{Introduction}

The duality between ${\cal N}=4$ Yang-Mills and string theory on 
$AdS_5 \times S^5$ is  by far the most well studied example of 
the Maldacena correspondence \cite{Maldacena:1997re,Aharony:1999ti}. 
 Another well studied  and interesting 
example of the correspondence is the case of 
the   duality between
type IIB string theory on $AdS_3\times S^3\times T^4$ and 
the ${\cal N}=(4,4)$ superconformal field theory  on a resolution of 
the symmetric product \cite{Maldacena:1997re,Maldacena:1998bw,Seiberg:1999xz}
\bej{defmani}
{\cal M} = T^{N}/S(N).
\eej
$AdS_3\times S^3\times T^4$ arises as a near horizon limit of the system
of $Q_1$ D1-branes and $Q_5$ D5-branes wrapped on $T^4$, then 
$N$   in \eq{defmani} is given by $Q_1Q_5$.
The duality sates that the spectrum of operators in the 
${\cal N}=(4,4)$ superconformal field theory on ${\cal M}$ should be the same
as the spectrum of type IIB string states in $AdS_3\times S^3\times T^4$. 
Operators which have large charges in the CFT should be dual
to classical string configurations \cite{Gubser:2002tv}. 

In this paper we consider operators with large $J$ charges, here
$J=J^3 + \tilde J^3$ the sum of the left and right $SU(2)$ R-charges of the
${\cal N}=(4,4)$ conformal field theory. 
We study states with finite $\Delta -J$, where $\Delta =L_0 + \tilde L_0$ the 
left and right conformal weights of the operators. 
Operators with $\Delta -J =0$ are chiral primaries which are the ground
states of the $\ZZZ_J$ twisted sector with $J^3$ charges
$(\frac{J-1}{2}, \frac{J-1}{2})$. We can then consider 
finite number of excitations of the following form on the chiral primary
\bej{intdefext}
J_{p_1}^{-} J_{p_1}^- \cdots J_{p_j}^- |0\rangle \otimes |0\rangle,
\eej
the vacuum in \eq{intdefext} denotes the
$\ZZZ_{J}$ twisted sector. 
 $J^-_{p}$ are operators which lower the left $J^3$ quantum number
and carry momentum $p$ in the $\ZZZ_J$ twisted sector,  
under the action of an element of $\ZZZ_J$,  $J^-_p|0\rangle$ picks up a
phase proportional to integer multiples of $p$. 
We can therefore  think of the insertions of $J_{p}^-$ as magnons or
impurities  that move
with momentum $p$

States of the form given in \eq{intdefext} were studied earlier in the limit
of small momentum $p$ and in first order in the $\ZZZ_2$ blow up
mode \cite{Lunin:2002fw,Gomis:2002qi,Gava:2002xb}. 
The dispersion relation of a single magnon was shown to be
\bej{intdisp}
\Delta -J = 1 +\frac{1}{2 \pi^2} \lambda^2 (Q_1Q_5) \frac{p^2}{4},
\eej
where
$\lambda$  is the coupling of the relevant $\ZZZ_2$ blow 
up mode in the 
symmetric product. 
In this paper we are interested in studying the magnons in the ``giant magnon''
 limit given by
\beaj{giantmag}
J\rightarrow\infty, &\qquad&  \tilde\lambda=  \lambda^2 (Q_1Q_5) = {\hbox{fixed}}, \\ \nonumber
p = {\hbox{fixed}}, &\qquad&  \Delta -J = {\hbox{fixed}}.
\eeaj
This differs from the plane wave limit \cite{Berenstein:2002jq}
 where $\tilde\lambda$
is infinite and it is $n = pJ$ which is kept fixed. 

Examining earlier  studies of the magnons  within perturbation theory in 
$\lambda$  and  the plane wave limit we argue that the magnons are BPS states in a 
centrally extended $SU(1|1)\times SU(1|1)$ superalgebra, the extended algebra has
$2$ more additional central charges. The centrally extended algebra can be 
written as a ${\cal N}=2$ Poincar\'{e} superalgebra in 3-dimensions with 
a single central charge. The remaining central charges play the role of 
the $3$-momentum in the Poincar\'{e} superalgebra. 
We then construct a dynamic spin chain representation of the 
extended algebra which carries these additional central charges and 
derive the following dispersion relation for a single magnon with momentum $p$
\bej{dispmag}
\Delta -J = \sqrt{ 1 + f(\tilde\lambda) \sin^2 \frac{p }{2}}.
\eej
where $f(\tilde\lambda)$ is an undetermined function of the coupling
$\tilde\lambda$. 
On rewriting the extended $SU(1|1)\times SU(1|1)$ as a $2+1$ Poincar\'{e}
superalgebra the above dispersion relation can be viewed as
the relativistic dispersion relation of a massive BPS particle in the 
$2+1$ Poincar\'{e} superalgebra. 
The spin chain representation constructed is such that
these additional central charges  vanish on physical states when we
impose the momentum constraint,  the algebra then collapses to the 
usual algebra. From the perturbative result in \eq{intdisp} we see that 
\bej{weakconst}
f(\tilde\lambda) = \frac{\lambda^2 Q_1Q_5}{\pi^2}, \quad {\hbox{for}} \quad
\tilde\lambda << 1.
\eej

Equation \eq{dispmag} is a BPS relation and the magnons
in question have large $J$ charge, therefore  we should expect to the find 
them as classical solutions to  the string sigma model on $AdS_3\times S^3\times
T^4$. These solutions are identical to  the giant-magnon solutions found by  \cite{Hofman:2006xt} in $AdS_5\times S^5$. Since these solutions only require the  
subspace $R\times S^2$, they continue to be solutions \footnote{Recently giant magnons in $AdS_3\times S^3$ and 
related solutions were studied in \cite{Lee:2008sk}.}   in 
$AdS_3\times S^3\times T^4$ with Ramond-Ramond flux through the $S^3$.  
After the identification of the momentum of the magnons to the geometrical
angle \cite{Hofman:2006xt} in the classical solution we obtain the following dispersion relation
for the magnon
\bej{dispmags}
\Delta -J = \frac{R^2}{\pi\alpha'} |\sin\frac{p}{2} |,
\eej
where $R^2$ is the radius of $S^3$ given by $R^2/\alpha'  = g_6 \sqrt{ Q_1Q_1} $ and
$g_6$ is the $6d$ string coupling.  Following the logic of \cite{Hofman:2006xt}
we write  the giant-magnon solution in
Lin-Lunin-Maldacena  (LLM) \cite{Lin:2004nb} 
like coordinates for $AdS_3\times S^3$ \cite{Liu:2004ru}
and construct the Killing spinors of the geometry. 
From  the solution of the Killing spinors  and the 
stretched string like nature of the giant magnon in the LLM
geometry  we infer that 
the solution carries the required  additional central charges to  render it BPS
in the extended $SU(1|1)\times SU(1|1)$ algebra. 
The BPS condition then implies the following dispersion relation 
at strong coupling for a single magnon
\bej{disstrong}
\Delta -J = \sqrt{ 1 + g_6 ^2 \frac{Q_1Q_5}{\pi^2} \sin^2 \frac{p}{2} }.
\eej
Comparison with  \eq{dispmag} we see that the 
\bej{strongconst}
f(\tilde\lambda)   = \frac{g_6^2 Q_1 Q_5 }{\pi^2},    \qquad {\hbox{for}}
\qquad  \tilde\lambda >>1
\eej
Thus identifying the coupling constant $\lambda = g_6$ and examining the 
weak coupling result in \eq{weakconst}, perhaps we can guess that
\bej{spec}
f(\tilde\lambda) = \frac{ g_6^2 Q_1 Q_5 }{\pi^2}, 
\eej
at all values of the coupling $\tilde\lambda$. Note that the 
dispersion relation in \eq{dispmag} also agrees with the plane wave limit
when the equality in \eq{spec} is satisfied 
\cite{Berenstein:2002jq,Lunin:2002fw,Gomis:2002qi,Gava:2002xb}
The dispersion relation is 
similar  to that of giant magnons in ${\cal N}=4$ Yang-Mills with $R^2$ in the 
dispersion relation replaced by the radius of $S^5$ instead of $S^3$. 

The organization of the paper is as follows: In the next section  we review the 
results of the analysis of magnons at small $p$ and small $\lambda$ in the 
symmetric product pointing out the evidence for the extended $SU(1|1)\times
SU(1|1)$ algebra. In section 3.  we write down the  extended $SU(1|1)\times
SU(1|1)$ algebra and show that it can be written as a ${\cal N}=2$ Poincar\'{e}
algebra, we then construct a dynamic spin chain representation of 
magnons  using this algebra which obeys the dispersion relation 
\eq{dispmag} and show that it is a BPS relation of the extended algebra.
 In section 4. we examine the magnons at strong coupling using  
 LLM coordinates for $AdS_3\times S^3$. We show that the magnons 
 carry the required central charges to be BPS in the extended $SU(1|1)\times SU(1|1)$
  algebra. Appendix A, B fill in the details necessary to show that the 
  giant magnon solution is supersymmetric. The method developed in 
  Appendix A enables one to determine the supersymmetries of 
   of a solution of IIB gravity with $S^1\times S^1\times T^4$ isometry  by embedding
   it as a solution of $(1,0)$ 6d gravity.

\section{Magnons in the symmetric product}

In this section we present a short review of the symmetric product conformal 
field theory. We then 
 specify the  magnon excitations  in the symmetric product 
whose conformal dimensions will be the subject of our interest and review the 
results of perturbation theory in $\lambda$. 

The boundary theory corresponding to the system of $Q_1$ number of 
D1-branes and $Q_5$ number of D5-branes in type IIB on $T^4$ is given 
by the ${\cal N}=(4,4)$ super conformal field theory on a resolution of symmetric 
product orbifold
\bej{deforbi}
{\cal M} = (T^4)^{Q_1Q_5}/S(Q_1Q_5).
\eej
The global part of the ${\cal N} =(4,4)$ algebra is given by the 
supergroup $SU(1,1|2)\times SU(1,1|2)$. The two copies arise from the
left movers and the right movers of the conformal field theory on 
${\cal M}$.  The bosonic part of the the supergroup $SU(1,1|2)$ 
consists of the global part of the conformal algebra $SL(2,R)$
whose generators are $L_0, L_\pm$ and 
the global part of the R-symmetry $SU(2)$ whose generators are
$J^3, J^\pm$. The  8 supercharges for $SU(1,1|2)$ are labeled by
$G_{1/2}^{ab}, G_{-1/2}^{ab}$, where $a \in\{+, -\}$ denotes the
quantum numbers of the charges under $SU(2)_R$ and $b \in\{ +, -\}$
denotes the quantum numbers of the charges under 
$SU(2)_I$  which is an outer automorphism of the ${\cal N}=(4,4)$
algebra. The subscript $\pm1/2$ refer to the weights of the charges
with respect to $L_0$. 
For our discussion the anti-commutation relations of relevance 
are
\beaj{globalg}
\{G_{-1/2}^{++}, G_{1/2}^{--} \} = 2(L_0  - J^3),  \qquad
\{G_{1/2}^{-+}, G_{-1/2}^{+-} \} = 2(L_0 - J^3).
\eeaj
From the above anti-commutation relations it is easy to see that 
the set of generators $\{G_{-1/2}^{++}, G_{1/2}^{--} ,  L_0 ,  J^3 \}$ or 
the set $\{G_{1/2}^{-+}, G_{-1/2}^{+-} , L_0 , J^3 \}$ each  form a 
$SU(1|1)$ sub-algebra with central charge $L_0-J^3$. 
Similarly there is an identical copy of the $SU(1,1|2)$ algebra from the 
right movers. We refer to these generators with a $\tilde{}$ superscript:
$\{\tilde L_0, \tilde L_\pm, \tilde J^3, \tilde J_{\pm}, \tilde G_{1/2}^{ab}, 
\tilde G_{-1/2}^{ab}\}$. To be specific,  and it 
will be justified by the subsequent discussion we will focus on the
$SU(1|1)\times SU(1|1)$ subalgebra generated by the following
\bej{firstsubalgbe}
\{ G^{+-}_{-1/2}, G^{-+}_{1/2}, (L_0- J^3) \}, \qquad
\{ \tilde G^{++}_{-1/2}, \tilde G^{--}_{1/2}, (\tilde L_0- \tilde J^3) \}.
\eej
The terms in the brackets $(L_0 -J^3)$ and $(\tilde L_0 -\tilde J^3)$
form the central charges of the $SU(1|1)\times SU(1|1)$ algebra.

Chiral primaries in the symmetric product CFT satisfy the conditions
$G_{1/2}^{ab}|\psi\rangle = \tilde G_{1/2}^{ab} |\psi\rangle = 
G_{-1/2}^{+b} |\psi\rangle = \tilde G_{-1/2}^{+b} |\psi\rangle =0$. 
They satisfy $L_0=J^3$ and $\tilde L_0 =\tilde J^3$
We will focus on chiral primaries which are the ground states in the 
$\ZZZ_J$ twisted sector with the left and the right $J^3$ charge given by
$(\frac{J-1}{2}, \frac{J-1}{2})$, we denote this chiral primary by $|0\rangle_n\otimes
|0\rangle_n$.
The construction of this chiral primary ground state using twist operators is 
given in \cite{Larsen:1999uk,David:1999ec}
\footnote{For a detailed review please see \cite{David:2002wn})}.  

We now consider the following excitations above this chiral primary
\beaj{defexcta}
|\phi_{p_1}\phi_{p_2}\cdots \phi_{p_j} \rangle_J\otimes |0\rangle_J = 
J^-_{p_1}J^-_{p_2} \cdots J^-_{p_j} |0\rangle_J\otimes|0\rangle_J.
\eeaj
where $J^-_{p}$ is given by 
\beaj{defjminus}
 J^{-}_{p} = \sum_{k=1}^J e^{ipk} J^-_{(k) }.
\eeaj
and $J^-_{(k)}$ is the lowering operator of the left moving $SU(2)$ R-current
of the $k$-th copy of the torus involved in the $\ZZZ_J$ twisted sector. 
 To satisfy orbifold group invariance condition we need to impose the condition
\bej{physcond}
\sum_i p_i =0.
\eej 
At the free orbifold point such a state is non-chiral in the left moving sector while
it is still chiral in the right-moving sector.  We now perturb the symmetric product 
CFT with the marginal operator constructed from the $\ZZZ_2$ twist field and 
which is a singlet component of $SU(2)_I$. This operator is dual to 
a combination of Ramond-Ramond 0-form and a 4-form on the dual gravity  type IIB
background. This perturbation is given by \cite{David:2002wn}
\bej{defpertb}
\lambda ( G_{-1/2}^{++}\tilde G_{-1/2}^{+-} - G_{-1/2}^{+-}\tilde G_{-1/2}^{++} )
\Sigma^{(1/2,1/2)}
+ \hbox{c.c},
\eej
where $\lambda$ is the coupling constant, and 
$\Sigma^{(1/2,1/2)}$ refers to the $\ZZZ_2$ twist operator of
charge $(1/2,1/2)$. Finally c.c refers to the expression 
involving the antichiral field $ \bar\Sigma^{(1/2,1/2)}$. 
On perturbing the CFT with this operator the excited states given in 
\eq{defexcta} are no longer right-chiral \cite{Gava:2002xb}
\footnote{We follow the reference \cite{Gava:2002xb},
related  work has been done in \cite{Lunin:2002fw,Gomis:2002qi,Hikida:2002in}.
The analysis of these build upon the detailed evaluation of the 
3-point functions of correlation functions of twist operators for
symmetric products  which were performed in 
\cite{Arutyunov:1997gi,Lunin:2001pw,Jevicki:1998bm}.}
 it picks up anomalous dimensions.

We now recall the  results of the evaluation of the anomalous dimensions
of the class of operators given in \eq{defexcta} 
to first order in $\lambda$ in the limit $J\rightarrow \infty$ and $p_i <<1$ together
with the number of excitations being small \cite{Gava:2002xb}.
Consider the state \footnote{In \cite{Gava:2002xb} the momentum of the state in \eq{prostate} is referred to by the label $n$, with  $2\pi n/J =  p$. }
\bej{prostate}
|\phi_p\rangle_J\otimes |0\rangle_J = J^-_{p} |0\rangle_J\otimes |0\rangle_J.
\eej
\begin{enumerate}
\item
To first order in $\lambda$ the action of $\tilde G^{-a}_{+1/2}$ flips the state
$|\phi_p\rangle_J\otimes |0\rangle_J$ from the $\ZZZ_J$ twisted sector to the
$\ZZZ_{J-1}$ twisted sector.
We write this as 
\beaj{shifsec}
\tilde G^{-a}_{1/2} |\phi_p\rangle_J\otimes |0\rangle_J \propto 
\epsilon^{ab}\lambda G^{+b}_{-1/2} |\phi_p\rangle_{J-1}\otimes|0\rangle_{J-1}.
\eeaj
The above transition clearly conserves the left and right $J^3, \tilde J^3$ charge.
\cite{Gava:2002xb} evaluated the following overlap in the limit $J\rightarrow \infty, p<<1$ to  first order in $\lambda$
\bej{overlap}
{}_{J-1}\langle 0| \otimes {}_{J-1}\langle \phi_p| G_{1/2}^{-a} \tilde G_{1/2}^{-b}
|\phi_p\rangle_J\otimes |0\rangle_J = \epsilon^{ab} \frac{\lambda p \sqrt{Q_1Q_5} }{4\pi}.
\eej
Note that from the equation in  \eq{shifsec} we see that the state $|\psi_p\rangle_J\otimes |0\rangle_J$  which was chiral on the right movers is no
longer chiral. 
\item
In \cite{Gava:2002xb} it was shown that to first order in $\lambda$ the following
commutation relation is obeyed on the state.
\bej{centralcom}
\{\tilde G_{1/2}^{-a}, G_{1/2}^{-b} \}|\phi_p\rangle_J\otimes |0\rangle_J = \epsilon^{ab} \lambda \int
dz\partial_z (z \bar z \bar\Sigma^{(1/2,1/2)} |\phi_p\rangle_J\otimes |0\rangle_J.
\eej
Note that the operator $\Sigma^{(1/2,1/2)}$ corresponds to the chiral
primary with charge $(1/2,1/2)$, therefore 
$\int dz\partial_z (z \bar z \bar \Sigma^{(1/2,1/2)})$ in \eq{centralcom} commutes with
the following set of generators 
$$\{\tilde G_{1/2}^{-a} , \tilde G_{-1/2}^{+a}, G_{1/2}^{-a}, G_{-1/2}^{+a}\}.$$
Thus with respect to the $SU(1|1)\times SU(1|1)$ subalgebra 
given in \eq{firstsubalgbe} the operator
$\int dz\partial_z( z\bar z \bar\Sigma^{(1/2,1/2)} )$ is central.
Furthermore the action of the operator evaluated at the leading 
order in $\lambda$ and for $p<<1$ is given by \cite{Gava:2002xb}
\bej{vanishphy}
\int dz\partial_z (z \bar z \bar \Sigma^{(1/2,1/2)})|\phi_p\rangle_J\otimes |0\rangle_J
\sim p |\phi_p\rangle_{J-1}\otimes |0\rangle_{J-1}.
\eej
Thus on physical states which satisfy \eq{physcond} the action of the
central element vanishes.
The commutation relation given in \eq{centralcom} can also be seen in the 
plane wave limit \cite{Gava:2002xb}. In the plane wave limit
of $AdS_3\times S^3$  it can be seen that \cite{Gava:2002xb} 
the charges obey the following commutation relations
\beaj{comppw}
\{G_{1/2}^{-a}, \tilde G_{1/2}^{-b} \} = \epsilon^{ab} \frac{1}{p^+} 
\sum_p p N_p, \\ \nonumber
\{G_{-1/2}^{+a}, \tilde G_{-1/2}^{+b} \}
= \epsilon^{ab} \frac{1}{p^+} \sum_p p N_p.
\eeaj
where $N_p$ is the oscillator number operator at momentum $p$ on the pp wave
and $p^+$ is the light cone momentum. From these commutation relations
also it is seen that 
on physical states, which satisfy the condition $\sum_p pN_p=0$, the anti-commutation
relations vanish.
\item
To the leading order in $\lambda$ the correction to $\Delta -J$ where 
$\Delta =L_0 +\tilde L_0$ and $ J = J^3+ \tilde J^3$ is given by
\cite{Gava:2002xb}
\bej{correctian}
\Delta -J = 1 +\frac{1}{2\pi^2}  \lambda^2 (Q_1 Q_5)\left( \frac{p^2}{4}\right).
\eej
\item
Note the that marginal deformation of the conformal field theory 
given in \eq{defpertb} is such that $L_0=\bar L_0$. Furthermore the
excited state given in \eq{prostate} is such that $L_0=\bar L_0$, therefore in 
perturbation theory it is clear that the 
change in conformal weights  of states is such that 
$\delta L_0 = \delta \bar L_0$
\end{enumerate}

In principle there could be the following transition from the $\ZZZ_J$ 
twisted sector to the $\ZZZ_{J+1}$ sector
\bej{2ndtrans}
 \tilde G^{+a}_{-1/2} |\phi_p\rangle_J\otimes |0\rangle_J \rightarrow
\epsilon^{ab} G^{-b}_{1/2}|\phi_p\rangle_{J+1}\otimes |0\rangle_{J+1}.
\eej
 Note that the $J^3, \tilde J^3$ charges are conserved under such transitions,
 but to first order in $\lambda$ such transitions are not present \cite{Gava:2002xb},

\section{The $SU(1|1)\times SU(1|1)$ dynamic spin chain model}

As we have seen  in the previous section that it is 
possible to obtain the conformal dimensions of the magnon excitations
within perturbation theory it is sufficient to restrict our attention to the 
action of the supercharges $\{G^{+-}_{-1/2}, G^{-+}_{1/2}, \tilde G^{+-}_{-1/2}
\tilde G^{-+}_{1/2}\}$ or 
the set of supercharges $\{G^{++}_{-1/2}, G^{--}_{1/2}, \tilde G^{++}_{-1/2}
\tilde G^{--}_{1/2}\}$.
From the commutation relations in \eq{globalg} it can be seen that
the above charges generate the 
subgroup $SU(1|1)\times SU(1|1)$ with central charges $L_0 - J^3$ and $\tilde L_0 -\tilde J^3$. 
To simplify the discussion we choose  one of the $SU(1|1)\times SU(1|1)$ 
algebra and define the generators as follows
\beaj{redefcharg}
G^{+-}_{-1/2} \rightarrow Q_1, &\qquad& 
\tilde G^{+-}_{-1/2} \rightarrow Q_2, \\ \nonumber
G^{-+}_{1/2} \rightarrow S_1, &\qquad &
\tilde G^{-+}_{1/2} \rightarrow S_2, \\ \nonumber
L_0 - J^3 \rightarrow  C_1, &\qquad& \tilde L_0 - \tilde J^3  \rightarrow C_2.
\eeaj
In terms of these variables, the $SU(1|1)\times SU(1|1)$ algebra is given by
\beaj{comrule1}
\{Q_1, S_1 \} = C_{1}, &\qquad& \{Q_2, S_2\} = C_{2} \\ \nonumber
\{Q_1, Q_2\} =0, &\qquad& \{ S_1, S_2\} =0, \\ \nonumber
\{Q_1, S_2\} =0, &\qquad& \{ S_1, Q_2\} =0.
\eeaj
 $C_{1}$ and $C_{2}$ are central 
elements of the algebra. 

The magnon excitations  given in \eq{defexcta} 
with momentum $p_i= 0$ belong to the 
BPS states of this algebra with $C_{1}= j, C_{2} =0$. 
We now consider magnons with momentum $p_i \neq 0$, these states
are not BPS in the above algebra as $C_{1}, C_{2} \neq 0$. 
But, on turning on interactions due to the marginal operator
in \eq{defpertb} we propose that the 
above algebra gets central extended with $2$ more additional
central charges. The magnons are then BPS states within this 
extended algebra and carry these central charges. 
 These central charges are such that on physical states 
they vanish. 
We then derive the dispersion relation relating the conformal dimensions
of the magnons to the momentum $p_i$.

\subsection{The extended $SU(1|1)\times SU(1|1)$ algebra}

From the anti-commutation relation \eq{centralcom} derived at first order in 
perturbation theory and the anti-commutation relations
\eq{comppw} obtained in the plane wave limit we see that the 
we should extend the $SU(1|1)\times SU(1|1)$ subalgebra such that
$\{ Q_1, Q_2\}$ and $\{ S_1, S_2 \}$ is non-trivial. Therefore
we consider the following central extension of the $SU(1|1)\times SU(1|1)$
algebra, given by the commutation  relations
\beaj{comrule2}
\{Q_1, S_1\} = C_{1} , &\qquad& \{Q_2, S_2\} = C_{2}, \\ \nonumber
\{Q_1, Q_2\} = C_3 - i C_4 , &\qquad&  \{S_1, S_2\} = C_3 + i C_4, \\ \nonumber
\{Q_1, S_2\} =0, &\qquad& \{ S_1, Q_2\} =0.
\eeaj
Note that we have extended the algebra by including $2$ more central charges
$C_3, C_2$,  further more since 
\bej{herprop}
Q_a^\dagger = S_a,
\eej
we have $\{Q_1, Q_2\}^\dagger = \{S_1, S_2\}$ the central charges in 
these two cases are related by a Hermitian conjugation. 
Note that the above central extension 
of $SU(1|1)\times SU(1|1)$ is different from that considered in 
\cite{Beisert:2007sk} which arises in certain sub-sectors of ${\cal N}=4$ Yang-Mills
\footnote{In the  extension considered by  \cite{Beisert:2007sk},
 the anti-commutators $\{Q_1, S_2\}$ and $ \{ S_1, Q_2\}$ were 
non-trivial.}

This central extension of the $SU(1|1)\times SU(1|1)$ in \eq{comrule2}
can be viewed as 
a  $N=2$ Poincar\'{e} superalgebra  in 3-dimensions with one central charge.
The remaining central charges play the role of $3$-momentum in the 
Poincar\'{e} superalgebra.  This is similar to the case of
giant magnons of ${\cal N}=4$ Yang-Mills, there the
magnons are BPS states of the  central extended 
 $SU(2|2)$ superalgebra  which can also be written as 
 a Poincar\'{e} superalgebra in 3-dimensions \cite{Hofman:2006xt}.
To  view the centrally extended 
algebra in \eq{comrule2}
as a ${\cal N}=2$ Poincar\'{e} algebra we first 
 define the following two component Majorana spinors in 3-dimension
\bej{defmaj}
q^1  = \left[ \begin{array}{c}
Q_1 + S_1 \\ i(Q_2 - S_2 )\end{array}\right], \qquad
q^2 = \left[ \begin{array}{c}
i ( Q_1 - S_1) \\ (Q_2 + S_2) 
\end{array} \right].
\eej
It is easy to see that from the property \eq{herprop} that these charges are real. 
We now can write the commutation relations for the extended algebra
in \eq{comrule2} as 
\bej{3dcomrule}
\{q_\alpha^i, q_\beta^j \} 
= 2 \delta^{ij} \hat p_\mu (\tilde\gamma^{\mu})_{\alpha\beta} + 2  \epsilon^{ij} \epsilon_{\alpha\beta} C_3.
\eej
Our conventions for the 3-dimensional  $\gamma$-matrices are as follows:
\bej{defgmat}
\gamma^0 = i \sigma^2 , \qquad \gamma^1 = \sigma^1, \qquad \gamma^2 = \sigma^3.
\eej
where $\sigma^i$ are Pauli matrices. We also define
\bej{defgmatt}
\tilde\gamma^\mu_{\alpha\beta} = (\gamma^\mu)^\gamma_\alpha \epsilon_{\gamma\beta}, \qquad \tilde \gamma^0 = - \delta^{\alpha\beta}, 
\qquad \tilde\gamma^1 = - \sigma^3, \qquad \tilde\gamma^2 =\sigma^1.
\eej
From \eq{comrule2} and the definition of $\tilde\gamma^\mu$ and the relation 
\eq{3dcomrule} we see that the momenta $\hat p_\mu$ are identified with the 
central charges as follows
\bej{cenident}
-\hat p_0- \hat p_1 = C_{1}, \qquad -\hat p_0 +\hat p_1 = C_{2}, \qquad \hat p_2 = C_4.
\eej
The algebra given in \eq{3dcomrule} is the super Poincare algebra in $3d$ with 
the central charge $C_3$.The remaining central charges of the extended
$SU(1|1)\times SU(1|1)$ algebra are identified with the momenta in $3d$
which commute with the supercharges. 
From the RHS of \eq{3dcomrule} we see that BPS states exist 
when 
\bej{bpscond}
\hat p_0^2 = \hat p_1^2 + \hat p_2^2 + C_3^2, \quad \hbox{or}\quad 
\frac{1}{4}( C_{1} + C_{2})^2 = \frac{1}{4} ( C_{1} - C_{2})^2 + C_3^2 + C_4^2.
\eej

\subsection{Dynamic spin chain representation}

We have seen that the 
magnon excitations given in \eq{defexcta} with $p_i=0$ are states
with $C_1 = j$ and $C_2=0$, one can also see that they satisfy the 
BPS condition \eq{bpscond}.  In this section 
following  \cite{Beisert:2005tm} we write down a representation of the extended 
$SU(1|1)\times SU(1|1)$ algebra in  terms of a dynamic spin chain which
carries the central charges $C_3, C_4$. 
We propose that these states correspond to magnons
with $p_i\neq 0$. 
The charges are turned on in such a way
that on physical states they vanish. 
We thus satisfy the property \eq{centralcom} and \eq{comppw} seen 
both in perturbation theory at first order  as well as in the plane wave limit. 
Using this representation of magnons we derive a dispersion relation
of the energy $C_1 + C_2 = \Delta -J =  (L_0 +\tilde L_0) -( J^3+ \tilde J^3)$ of 
these magnons.  This dispersion relation satisfies the
BPS condition \eq{bpscond} and thus is valid at all orders in 
interaction. In the next section we identify the magnons at strong coupling 
and show that they indeed satisfy the same dispersion relation.

The vacuum state of the spin chain is a chrial primary   denoted as 
\bej{vacsta}
|0\rangle_J\otimes |0\rangle_J = | \ldots \psi\psi\ldots \rangle\otimes
|\dots\tilde\psi\tilde \psi \ldots \rangle.
\eej
This state represents the chiral primary or weight $(\frac{J-1}{2} , \frac{J-1}{2})$
in the $\ZZZ_J$ twisted sector of the symmetric product.  We work in the limit
$J\rightarrow \infty$. It is convenient to think of each $|\psi\rangle$ as
state which carries weight $L_0=1/2$ and $J^3 =1/2$. In the language of the 
twist field it is a  $\ZZZ_2$ twist field which implements the permutation
between two copies of the torus $T^4$. 
Thus in the vacuum state in \eq{vacsta}, there are $J-1$ $|\psi\rangle$'s 
each carrying $L_0=1/2, J^3=1/2$ at the $J-1$ sites for the 
right moving vacuum and similarly there are $J-1$  $|\tilde \psi \rangle$
which carries weight $\tilde L_0 =1/2, \tilde J^3 =1/2$ at $J-1$ sites 
for the left moving vacuum. 
All charges $Q_1, Q_2, S_1, S_2$ annihilate the vacuum \eq{vacsta}  since 
it is a chiral primary. 
From now on we will work in the limit of the infinite $J\rightarrow \infty$ chain.
We consider the following excitations on this vacuum
\bej{excits}
|\phi_{p_1}\ldots \phi_{p_j}\rangle\otimes |0\rangle = 
\sum_{n_1,<<\ldots<< n_j} e^{ip_1n_1} \ldots e^{ip_jn_j}
| \cdots \psi\psi \cdots \phi_1 \cdots \phi_2 \cdots \phi_j \cdots \psi\psi \cdots \rangle
\otimes |0\rangle.
\eej
Note that we have removed the subscript $J$ from the kets since we are working
in the strict $J\rightarrow \infty$ limit. 
Here the state $|\phi\rangle$ represents a state with $L_0=1/2, J^3 =-1/2$ with 
It is obtained from the state $|\psi\rangle$  by the 
\bej{relphispi}
|\phi\rangle = J^- |\psi\rangle.
\eej
Thus  excitations given in \eq{excits} can
be  obtained   following action of $J^-_p$ on the 
vacuum
\bej{exciprod}
J_{p_1}^- J_{p_2}^- \cdots J_{p_j}^- |0\rangle\otimes |0\rangle,
\eej
where 
\bej{defj-}
J_{p}^-  = \sum_{l} e^{i p l }J^{-}_{(l)}.
\eej
$J^{-}_{(l)}$ acts on the state $|\psi\rangle$ at site $l$.
Thus the state in \eq{excits} corresponds to the state defined in 
\eq{defexcta}. 
The central charges at zeroth order in the coupling of the theory of this state
is given by $C_{1}= j, C_{2}=0$.

To define the action of the charges $Q_a, S_a$ on the general excited
state \eq{excits} we first define their action on the simple
state with one $\psi$ excited to $\phi$ on the extreme left.
Let
\bej{defsimexcit}
|\phi\rangle\otimes |0\rangle = |\phi \psi\psi\cdots \rangle\otimes
|\tilde\psi\tilde\psi \cdots\rangle,
\eej
then the action of the charges on this state is given by
\beaj{statrule}
Q_1 |\phi  \rangle \otimes |0\rangle = a |\psi^+\phi\rangle \otimes |0\rangle,
\\ \nonumber
Q_2|\phi \rangle \otimes |0\rangle=  a' |\phi\rangle\otimes |\tilde{\psi}^+\rangle,
\\ \nonumber
S_1|\phi\rangle \otimes |0\rangle = b |\psi^-\phi\rangle \otimes |0 \rangle,
\\ \nonumber
S_2 |\phi \rangle \otimes |0\rangle = b' |\phi\rangle \otimes |\tilde{\psi}^-\rangle.
\eeaj

In the above equations the presence of $\psi^+$ refers to the fact that that 
state is in the $\ZZZ_{J+1} \times \ZZZ_J$ twisted sector, 
while the presence of $\psi^-$ refers to the 
fact that the state is in the $\ZZZ_{J-1}\times \ZZZ_J$
twisted sector. 
Similarly the presence of
 $\tilde\psi^+$ refers to the  fact that that the state is in the 
 $\ZZZ_J\otimes \ZZZ_{J+1}$ twisted sector, and the presence of 
 $\tilde\psi^-$ refers to the fact that the state is in the 
 $\ZZZ_J\otimes \ZZZ_{J-1}$ sector.  $a, a', b,b'$ are constants which depend on the 
interaction strength of the theory and should vanish at the zeroth 
order in coupling. These transition rules are motivated 
from the observations given in \eq{shifsec} and  \eq{overlap}
seen in the symmetric product conformal field theory at first order in 
$\lambda$.  
In the extended algebra  \eq{comrule2} we have $Q_1^2 =0, S_1^2=0, Q_2^2=0, 
S_2^2=0$.
We write this as
\beaj{trivident1}
Q_1|\psi^+\phi\rangle \otimes |0\rangle =0, \\ \nonumber
Q_2|\phi\rangle\otimes |\tilde\psi^+\rangle =0, \\ \nonumber
S_1|\psi^-\phi\rangle \otimes |0 \rangle =0, \\ \nonumber
S_2 |\phi\rangle \otimes |\psi^-\rangle =0.
\eeaj

From the above rules it is clear that the the difference of the twists in the between
the left and the right moving sectors can be at the most $\pm 1$.
To impose the anti-commutation relations $\{Q_1, S_2\}=0, \{Q_2, S_2\} =0$ on 
the states of spin chain we assume the following states in the spin chain 
are proportional.

\beaj{equivalence}
Q_1|\Psi\rangle  =  g S_2| \Psi\rangle, \qquad
Q_2|\Psi\rangle = g' S_1|\Psi\rangle.
\eeaj

where $|\Psi\rangle$ is any state obtained by the action of charges 
on the state $|\phi\rangle \otimes |0\rangle$. The above equation is
motivated by the observation \eq{shifsec} seen in perturbation theory.
It is clear that using the above equation and the fact $Q_1^2 =0, S_1^2=0, Q_2^2=0, 
S_2^2=0$ the anti-commutation relations $\{Q_1, S_2\}=0, \{Q_2, S_2\} =0$
on the excited states are seen to hold.
Now the only other non-trivial sequences of action of charges to specify 
the representation  are
$Q_1 Q_2, Q_2 Q_1, S_1 S_2, S_2S_1, Q_1S_1, S_1Q_1, Q_2S_2, S_2Q_2$. 
 We write these 
\beaj{trivident2}
Q_1Q_2 |\phi\rangle\otimes |0\rangle
 &= &a a'|\psi^+\rangle\otimes|\tilde\psi^+\rangle,
\\ \nonumber
Q_2Q_1|\phi\rangle\otimes |0\rangle &=&
 aa' \gamma |\psi^+\rangle\otimes|\tilde\psi^+\rangle,
\\ \nonumber
S_1S_2| \phi\rangle\otimes| 0\rangle &=&
bb' |\psi^-\phi\rangle \otimes |\tilde\psi^-\rangle, \\ 
\nonumber
S_2S_1| \phi\rangle\otimes |0\rangle &=&
bb'\gamma' |\psi^-\phi\rangle \otimes |\tilde\psi^-\rangle, \\ \nonumber
S_1Q_1 |\phi\rangle\otimes |0\rangle &=& ab |\phi\rangle\otimes |0\rangle, 
\\ \nonumber
Q_1 S_1 |\phi\rangle\otimes |0\rangle &=& ab\tilde\gamma |\phi\rangle\otimes |0\rangle, 
\\ \nonumber
S_2Q_2 |\phi\rangle\otimes |0\rangle &=& a'b' |\phi\rangle\otimes| 0\rangle, 
\\ \nonumber
Q_2S_2 |\phi\rangle\otimes| 0\rangle &=& a'b'\tilde{\gamma }^{\prime}|\phi\rangle\otimes |0\rangle. 
\eeaj
Note  that the action of the $Q_1Q_2$ does not anti-commute with the 
action of $Q_2Q_1$ for $\gamma \neq -1$ this is precisely what we 
require if the central charges $C_3, C_4$ need to be turned on.
A similar statement holds for the action of $S_1S_2$ and $S_2S_1$.
Thus interchanging $Q_1Q_2$ picks up a factor of $\gamma$ and 
interchanging $S_1S_2$ picks up a factor of $\gamma'$. 
Similarly note that $Q_1S_1$ and $Q_2S_2$ also do not anti-commute.
 
Now using the defintion of the spin chain representation given in 
\eq{statrule}, \eq{trivident1} and \eq{trivident2} we read out the
central charges carried by the representation. Examining 
 the relations $\{Q_1, S_1\}$
and $\{Q_2, S_2\}$  on the excited state in \eq{defsimexcit} give
\beaj{defc11}
C_{1} |\phi\rangle \otimes |0\rangle & =& ab(1+\tilde \gamma)  
|\phi\rangle \otimes |0\rangle ,
\\ \nonumber
C_{2} |\phi\rangle \otimes |0\rangle &=& a'b'(1+\tilde \gamma')  |\phi\rangle \otimes |0\rangle. 
\eeaj
Finally from the relations $\{Q_1, Q_2\}$ and $\{S_1, S_2\}$ we have
\beaj{defp1p2}
(C_3 -iC_4 ) |\phi\rangle \otimes |0\rangle  =  
(1+\gamma) aa'|\psi^+\phi\rangle \otimes
|\tilde\psi^+\rangle, \\ \nonumber
(C_3 +iC_4 ) |\phi\rangle \otimes |0\rangle =  
(1+ \gamma') bb'|\psi^-\phi\rangle\otimes
|\tilde\psi^-\rangle.
\eeaj
In deriving this we used 
the rules given in \eq{trivident2}
Note that from the action of the charges given in \eq{statrule},
\eq{equivalence},  \eq{trivident1} and 
\eq{trivident2} it 
can be shown that the charges $C_3, C_4$ are central.
Therefore the rules \eq{statrule} , \eq{equivalence},
\eq{trivident1}  and \eq{trivident2}
 define a representation
of the extended $SU(1|1)\times SU(1|1)$ algebra.

We are interested in representation such that 
the  central charges $C_3, C_4$ are turned on locally in a state but globally on 
physical states these central charges vanish. For this we  consider an excitation
of definite  momentum $p$ given by 
\bej{sinexcip}
|\phi_p \rangle\otimes |0\rangle 
= \sum_k e^{ipk} | \ldots \psi\psi \ldots \phi \ldots \psi\psi \ldots \rangle \otimes
|0\rangle.
\eej
Insertion or removal of  $\psi$  to the immediate left  of  the excitation we obtain
\beaj{relphase}
|\psi^{\pm} \phi_p\rangle \otimes |0\rangle
=\sum_k e^{ipk}|\ldots  \psi^\pm\phi \ldots \rangle\otimes |0\rangle, \\ \nonumber
= \sum_k e^{ipk \mp  ip} |\ldots\psi\psi \ldots \phi \ldots \rangle \otimes |0\rangle.
\eeaj
Thus upto a phase we can shift the insertion or removal of $\psi$ to the very end.
We therefore have the relation
\bej{phaserelation}
|\psi^\pm \phi_p\rangle\otimes |0\rangle
 = e^{\mp ip} |\phi \psi^\pm\rangle\otimes|0\rangle.
\eej
Thus the action of the central charge $C_3-iC_4$ on the tensor product of excitations
is given by
\beaj{actenccrg}
(C_3- iC_4) | \phi_{p_1}\ldots \phi_{p_j} \rangle\otimes |0\rangle
= {\cal C}  | \phi_{p_1}
\ldots \phi_{p_j} \psi^+\rangle\otimes |\psi^+\rangle, \\ \nonumber
{\cal C}  = \sum_{k=1}^j a_ka_k' (1+ \gamma) \prod_{l =k+1}^j e^{-ip_l}.
\eeaj
Therefore ${\cal C}$ should vanish on physical states, this is obtained by setting
\bej{valaa}
a_ka_k'(1+\gamma) = \alpha\left( e^{-ip_k}-1\right), 
\eej
where $\alpha$ is a function of the coupling $\tilde\lambda$,
the normalization of $\alpha$ above is for convenience.
With this condition it is easy to see that ${\cal C}$ is given by
\bej{valp}
{\cal C} = \alpha\sum_{k=1}^j ( e^{-ip_k} -1) \prod_{l=k+1}^j e^{-ip_l} 
= \alpha  \left( \prod_{k=1}^j e^{-ip_k} -1 \right).
\eej
Thus on physical state ${\cal C} $ vanishes. Similarly we see that we should set 
\bej{valbb}
b_kb_k'(1+\gamma')= \alpha^*
\left(e^{ip_k} -1\right),
\eej
which ensures  $C_3+ iC_4$ also vanishes on physical state.
Note that the central charge $C_3  + i C_4$ is the complex conjugate of 
$C_3-i C_4$ which has been 
implemented in \eq{valaa} \eq{valbb}. 
For the state $|\phi_p\rangle\otimes|0\rangle$ we have $C_{1}-C_{2}=1$, 
since
\bej{manip1} 
C_{1}-C_{2}=(L_0 - J^3)  - (\tilde L - \tilde J^3) = \tilde J^3 -J^3.
\eej
Here the last equality  follows due to property 4. seen in perturbation theory,
that is the condition $L_0 =\tilde L_0$ is maintained in perturbation theory.
from \eq{defc11} we have
\bej{condc11}
(1+\tilde\gamma) ab- (1+\tilde\gamma') a'b' =1.
\eej
Then from the equations \eq{valaa}, \eq{valbb} and \eq{condc11} we obtain
\beaj{andim}
\Delta -J = C_{1} + C_{2} &=& \left( (1+\tilde\gamma) ab + (1+\tilde\gamma')a'b' \right) \\ \nonumber
 &=& \sqrt{1 + 16\alpha^*\alpha \frac{(1+\tilde\gamma)( 1+\tilde\gamma')}{ (1+ \gamma)(1+\gamma')}\sin^2(\frac{p}{2}) }, \\ \nonumber
 &=& \sqrt{ 1+ f(\tilde\lambda) \sin^2(\frac{p}{2}) }.
\eeaj
We thus have obtained the dispersion relations for the magnons.
Note that the above relation satisfies the  BPS condition given in 
\eq{bpscond}.
From the comparison of the correction to $\Delta -J$ computed 
at weak coupling  and small mometum given in \eq{correctian} we see that
\bej{coupident1}
f(\tilde\lambda)=  \lambda^2 \frac{Q_1Q_5}{\pi^2}.
\eej
For the tensor product exicitation given in \eq{excits}  with the assumption that the
exciations are well separated we obtain the dispersion relation
\bej{tendisprel}
\Delta-J  = \sum_{ i = 1}^j \sqrt{ 1 + f(\tilde\lambda) \sin^2(\frac{p_i}{2}) }.
\eej

\sectiono{Magnons at strong coupling} 
\label{strong}

 The classical solutions of the string sigma model
found  by \cite{Hofman:2006xt} for the case of 
$AdS_5\times S^5$ continue to be solutions in $AdS_3\times S^3\times T^4$. This is 
because they require only the subspace $R\times S^2$ which is also available 
in $AdS^3\times S^3$. It is only when there is   Ramond-Ramond flux
through the $S^3$ the equations of motion discussed by \cite{Hofman:2006xt} continue
to be the same for the case of $AdS^3\times S^3$. Thus the discussion in this 
section applies for pure Ramond-Ramond flux through the $S^3$.
 We first start with a short review of the magnon solution and obtain the 
 dispersion relation of the magnons at strong coupling. 
 We then provide a detailed analysis of the supersymmetry preserved by 
 these magnons by following the logic outlined in \cite{Hofman:2006xt}.
 This involves writing the solution in LLM like coordinates 
 for the case of $AdS_3\times S^3$ in which the magnon solution is just 
 a stretched string. By studying the Killing spinors and a particular one-form
 which corresponds to the gauge transformation of the NS B-form under the 
 action of two supersymmetries we see that the magnon solution carries 
 the required central charges to be BPS. 
 The reason it is BPS is the same as the reason a stretched string 
 is BPS in flat space, in fact the supersymmetry algebra turns out to be the 
 the extended $SU(1|1)\times SU(1|1)$ algebra discussed in 
  in section 2.

\subsection{Magnon dispersion relation at strong coupling}

The near horizon geometry of the D1-D5 system for large $\tilde \lambda$ is 
$AdS_3\times S^3\times T^4$ 
described by the following metric \cite{Maldacena:1998bw}:
\be \label{metric}
ds^{2}=R^{2}(-\cosh^2\rho dt^{2}+d\rho^{2}+\sinh^2\rho d\varphi^{2}+d\Omega_{3}^{2}
) + ds^2([T^4]),
\ee
where
$d\Omega_{3}^{2}$ is the metric on the unit three sphere given by 
\be \label{sphere}
d\Omega_{3}^{2}=d\theta^{2}+\sin^{2}\theta d\phi^{2}+\cos^{2}\theta d\psi^{2}, 
\ee
with $0\leq \theta\leq \pi $, $0\leq \phi\leq 2\pi$, $0\leq \psi \leq 2\pi$, 
and $ ds^2([T^4]$ is the flat metric on the four torus given by
\bej{4torus}
ds^2[T^4] =  \alpha' \sqrt{\frac{Q_1}{vQ_5}} \left( 
dx_6^2 + dx_7^2 + dx_8^2 + dx_9^2 \right).
\eej
$v$ is the volume of asymptotic volume of the four torus in string units, 
and 
\bej{defsigrad}
R^{2}= \alpha' g_6 \sqrt{Q_{1}Q_{5} },
\eej
 with $g_6 = g_s /\sqrt{v}$.
Note that we have used global coordinates to describe $AdS_3$. 
We identify $\phi$ to be the coordinate 
conjugate to angular momentum $J =J^3+\tilde J^3$. 
The string ground state with $E-J=0$ corresponds to a lightlike trajectory that moves along $\phi$ with $\phi - t=$ constant and sits at $\theta=\pi/2$ . 
and at the center of $AdS_{3}$, $\rho =0$.
 Now to obtain the string configuration that corresponds to a solution carrying momentum $p$ with the least amount of energy $\epsilon=E-J$ we follow \cite{Hofman:2006xt}. We first choose the point $\psi=0$ on the  circle $S^{1}$ parameterized by $\psi$. This 
 point along with $\theta$ and $\phi$ form a $S^{2}$. After we include time, the motion takes place in $R \times S^{2}$ where $R$ is parameterized by the time coordinate. We now write the Nambu-Goto action for the string in this background by choosing the world sheet coordinates to be:
\be \label{WS}
t=\tau, \quad\quad \phi - t =\sigma,
\ee 
and we consider a configuration where $\theta$ is independent of $\tau$. The Nambu-Goto action takes the form:
\be \label{nambu}
S=\frac{1}{2\pi\alpha^{\prime}}\int{d\tau d\sigma \sqrt{-det \gamma}},
\ee
where $\gamma$ is the induced metric on the world sheet given by 
\be \label{gamma}
\gamma_{ab}=g_{\mu\nu}\frac{\partial X^{\mu}}{\partial\sigma^{a}}\frac{\partial X^{\nu}}{\partial\sigma^{b}},
\ee
where $a,b=0,1$ and $\sigma^{0}=\tau$ and $\sigma^{1}=\sigma$. After taking into account the worldsheet parameterization given in \eq{WS} we get the following action,
\be \label{Nambu}
S=\frac{R^{2}}{2\pi\alpha^{\prime}}\int{d\tau d\sigma \sqrt{\cos^{2}\theta{\theta^{\prime}}^{2}+\sin^{2}\theta}},
\ee
here $\theta'$ refers to derivative with respect to $\sigma$. 
Following \cite{Hofman:2006xt} we integrate the equations of motion and get
\be \label{eom}
\sin\theta=\frac{\sin\theta_{0}}{\cos\sigma}, \quad\quad -\left(\frac{\pi}{2}-\theta_{0}\right)\leq\sigma \leq\frac{\pi}{2}-\theta_{0},
\ee
where $0\leq\theta_{0}\leq\frac{\pi}{2}$ is an integration constant.
The difference in angle between the two endpoints of the string at 
a given time
$\delta\sigma =2\left(\frac{\pi}{2}-\theta_{0}\right)$ is identified with the momentum $p$ of the magnon \cite{Hofman:2006xt},
 we write this as
\bej{basident}
\delta\phi=2\left(\frac{\pi}{2}-\theta_{0}\right) =p,
\eej
 and one also obtains the energy $E-J$ which is the Noether charge 
corresponding to shifts in $\sigma$ to obtain
\bej{noether}
E-J= \frac{R^2}{\pi \alpha'} \cos \theta_0 = 
\frac{R^{2}}{\pi\alpha^{\prime}}\sin\frac{\delta\phi}{2}.
\eej
After we have identified $\delta\phi$ with $p$, we obtain the  following strong coupling result of the dispersion relation,
\be \label{dispersion}
E-J=\frac{R^{2}}{\pi\alpha^{\prime}}\left|\sin\frac{p}{2}\right|,
\ee
Note that this dispersion relation agrees with the strong coupling limit
of \eq{dispmag} if $f(\tilde\lambda) \rightarrow g_6 ^2 Q_1Q_5/\pi^2$ for
$\tilde \lambda \rightarrow \infty$. 
We now proceed to demonstrate that these magnons are supersymmetric.

\subsection{Supersymmetry preserved by magnons}
 \label{s1}

There are two crucial ingredients to demonstrate that these magnons are BPS solutions
of type IIB on $AdS_3\times S^3\times T^3$. The first one is to demonstrate that
in a particular coordinate system  the magnon solution is just a straight stretched string.
For this 
 we  write the 
solution given in  \eq{eom} using the LLM \cite{Lin:2004nb} coordinates suitable for
$AdS_3\times S^3$.  
The $AdS_3\times S^3$ metric in these coordinates is given by \cite{Liu:2004ru}
\beaj{ads3solm}
ds_6^2 = -h^2 ( dt + V_i dx^i)^2 
+ h^2( dy^2 + \delta_{ij} dx^idy^j) 
+ y ( e^G d\Omega_1^2 + e^{-G} d\tilde\Omega_1^2), \\
\eeaj
where
\bej{defhzv}
h^{-2} = 2y \cosh G, \qquad z \equiv \frac{1}{2} \tanh G, \qquad
dV = -\frac{1}{y} *_3 dz.
\eej
and $z$ satisfies the following equations
\beaj{zequ}
\left( \partial_i^2  + y\partial_y \frac{1}{y} \partial_y \right) z =0, \\ \nonumber
\partial_i z\partial_i z + \partial_y z\partial_y z = 
\frac{ ( 1-4z^2)^2}{4y^2}.
\eeaj
For the $AdS_3\times S^3$  metric $z$ is such that in the plane $y=0$, 
$z=1/2$ is a circular region of radius $R$. The above metric is a fibration of
the time direction $t$ and the two $S^1$'s denoted by $\Omega_1$ and $\tilde\Omega_1$
over the three dimensional space characterized by $x_1, x_2, y$.
We can obtain  the conventional global
coordinates of $AdS_3\times S^3$ given in \eq{metric}  using the following
change of coordinates
\beaj{chgcoord}
& y = \sinh\rho \cos\theta, \qquad
r = \cosh\rho \sin \theta, \qquad
\sigma = \phi - t, \\ \nonumber
& x_1 = r\cos\sigma, \qquad x_2 = r\sin\sigma,  \\ \nonumber
& \varphi =\Omega_1, \qquad  \psi = \tilde\Omega_1.
\eeaj
Using this change of variables, the metric on the plane $y=0$
for $r<1$  is of the form
\bej{planemet}
ds^2 = R^2\left[
-(1-r^2) \left( dt - \frac{r^2}{1-r^2} d\sigma \right)^2 + 
\frac{ dr^2 + r^2 d\sigma^2}{1 - r^2} + ( 1 -r^2) d\psi^2 \right].
\eej
We now repeat the analysis of \cite{Hofman:2006xt} for this case. 
From \eq{chgcoord} we see that $r^2 = \sin^2\theta = x_1^2 + x_2^2$, since
$\rho =0$,  the solution \eq{eom} can be written as
\bej{stline}
r\cos\sigma = x_1 = \cos\theta_0 = {\hbox{constant}}.
\eej
Thus the magnon solution is just a straight stretched string in these coordinates. 
The string is stretched between two points on a circle. 
Note that the energy $E -J$ of the magnon is just the length of the string with flat metric
on  $(x_1, x_2)$ plane, 
\bej{rrwrnoeth}
E-J = \frac{R^2}{\alpha'\pi} \Delta x_2 = \frac{R^2}{\alpha'\pi}\cos\theta_0.
\eej
Finally in these coordinates the angle subtended by the string at the centre of the 
circle is related to $p$ by $2\theta_0 = \pi -p$ from \eq{basident}

The second ingredient 
needed to show that the  giant magnon solution preserves supersymmetry is to
 demonstrate the
it carries the appropriate central charges. 
These central charges arise
due to the fact that it is just a straight stretched string in the LLM like coordinates and thus it has the appropriate winding charge needed to make it supersymmetric.
This is the same reason that stretched strings in flat space are BPS. 
To show that that the giant magnon solution carries these charges 
we again follow the logic outlined in  \cite{Hofman:2006xt}. 
We first need to find out the Killing spinors  for the $AdS_3\times S^3$ solution
in LLM like coordinates. 
We start with the LLM like ansatz for type IIB with $S^1\times S^1$ isometry,
this ansatz accommodates $AdS_3\times S^3\times T^4$ as a solution 
\cite{Liu:2004ru}.  
 In the near horizon geometry 
of the D1-D5 system the Ramond-Ramond 3-form  
 is self dual, therefore
 we take $D=10$, IIB  supergravity with the 
 following bosonic  fields turned on: the metric $G_{MN}$, the 2-form potential
 $ C^{+}_{MN}$   with self dual field strength i.e $F_{(3)}=*F_{(3)}$ where $F_{3}=dC^{+}_{(2)}$.  Let 
 $\psi_{M}$ be  the gravitino which is a right handed Weyl spinor i.e it obeys the condition $\Gamma_{11}\psi_{M}=\psi_{M}$, $\lambda$ is the dilatino which is also a right handed Weyl spinor. Notice that  we have set the Ramond-Ramond 4-form potential and axion-dilaton to zero. We specify a reduction of the form :
\ben \label{a1}
ds_{10}^{2} &=& g_{\mu \nu}dx^{\mu}dx^{\nu}+e^{H(x)+G(x)}d\phi^{2}+e^{H(x)-G(x)}d{\tilde \phi}^{2}+ dx_{s}dx^{s}, \nonumber \\
F_{(3)} &=& -\frac{1}{2}F_{(2)} \wedge d\phi -\frac{1}{2}\tilde{F}_{(2)} \wedge d\tilde{\phi},
\een
where
\be \label{a2}
F_{(2)}=F_{\mu \nu}dx^{\mu} \wedge dx^{\nu}, 
\ee
 $s = 6, 7, 8, 9$ the  directions along $T^4$ and 
 $\mu = 0, 1, 2, 3$. The above ansatz preserves the $SO(2) \times SO(2)$ isometry.
This ansatz corresponds to setting the gauge fields from the components of the 
metric and the 2-form potential
: $g_{\mu \phi}$ and $C_{\mu \tilde{\phi}}$  and the scalars: 
 $C^{+}_{\phi \tilde{\phi}}$ and $g_{\phi \tilde{\phi}}$ components) to zero. This is an inconsistent truncation of the theory but as argued in \cite{Liu:2004ru}, this inconsistency manifests itself in one additional constraint  given in the second line of \eq{zequ}.
Since we have set the axion-dilaton 
to a constant and the Ramond-Ramond 4-form potential to zero, the supersymmetry variation of dilatino and gravitino takes the following form \cite{Liu:2004ru,Schwarz:1983qr}.
\ben \label{a3}
\delta \psi_{M}&=&\nabla_{M} \varepsilon -\frac{1}{96}(\Gamma_{M}\Gamma^{NPQ}+2\Gamma^{NPQ}\Gamma_{M})F_{NPQ}\varepsilon^{*}, \nonumber \\
\delta \lambda&=& -\frac{i}{24}F_{MNP}\Gamma^{MNP}\varepsilon.
\een
We use the convention that 10d gamma matrices are purely imaginary, explictly
they are given in \eq{newgamma}. 
Setting the dilatino variation to zero gives the 6d chirality condition on the spinor
\bej{6dchir0}
\Gamma^0\Gamma^1\Gamma^2\Gamma^3\Gamma^4\Gamma^5\epsilon =
-\epsilon.
\eej
Since the spinor $\epsilon$ is a 10 Weyl spinor we also have the following condition
\bej{t4cond0}
\Gamma^6\Gamma^7\Gamma^8\Gamma^9 \epsilon =-\epsilon.
\eej
The gravitino variation in the $0, 1, 2, 3, \phi, \tilde\phi $ directions are given by
\beaj{grav}
\delta\psi_\mu &=& \nabla_\mu\epsilon -\frac{1}{16} \gamma_{\rho\sigma}F^{\rho\sigma}e^{-\frac{1}{2}(H-G)}
\gamma_\mu \tilde{\varepsilon}\hat{\sigma_1} \epsilon^*, \nonumber \\
\delta\Omega_H &=& \frac{1}{2}( \partial_\mu H\gamma^\mu\tilde{\varepsilon}\hat{\sigma_1})\epsilon + e^{\frac{H+G}{2}} \partial_\phi\epsilon - ie^{-\frac{H-G}{2}} \gamma^5
\partial_{\tilde\phi} \epsilon, \\ \nonumber
\delta\Omega_G &=& \frac{1}{2}( \partial_\mu G \gamma^\mu\tilde{\varepsilon}\hat{\sigma_1})\epsilon+ e^{-\frac{H+G}{2}} \partial_\phi \epsilon +  ie^{-\frac{H-G}{2}} \gamma^{5}\partial_{\tilde\phi} \epsilon  -\frac{1}{8} \gamma_{\rho\sigma} F^{\rho\sigma}\epsilon^* .
\eeaj
Here $\delta \Omega_H$ and $\delta\Omega_G$ are defined as
\bej{defdOmeg}
\delta\Omega_H = \delta\psi_\phi - e^G\Gamma_5\Gamma_4\delta\psi_{\tilde\phi}, 
\qquad 
\delta\Omega_G = \delta\psi_\phi + e^G\Gamma_5\Gamma_4\delta\psi_{\tilde\phi}.
\eej
Note that these gravitino variations are different from that obtained
by \cite{Liu:2004ru} for the case of 6d $(1, 0)$ gravity. 
Due to the the occurrence of 
$\epsilon^*$ on the RHS of the variations the solutions of the Killing spinors obtained
by \cite{Liu:2004ru} for the case of 6d $(1,0)$ gravity cannot be directly 
embedded in IIB gravity. 
But as shown in appendix \ref{sugra}.  for the following   spinor
\bej{ccspin}
\tilde\epsilon = \epsilon_R +  i (1\otimes\tilde\varepsilon\otimes\hat\sigma_1) \epsilon_R.
\eej
the gravitino variations reduce to that obtained by \cite{Liu:2004ru}. 
In \eq{ccspin}  subscripts stand for the real part of the spinor $\epsilon$. 
Rewriting the gravitino variation \eq{grav} in terms of the spinor   $\tilde\epsilon$  from, we get  the following equations
\beaj{gravmod}
\delta\tilde\psi_\mu &=& \nabla_\mu \tilde\epsilon -\frac{i}{16}e^{-\frac{1}{2}(H-G)} \gamma_{\rho\sigma}
F^{\rho\sigma}\gamma_\mu \tilde\epsilon \\ \nonumber
\delta\tilde\Omega_H &=& -i\frac{1}{2} \partial_\mu H \gamma^\mu \tilde\epsilon
+ e^{-\frac{H+G}{2}} \partial_\phi \tilde\epsilon - ie^{-\frac{H-G}{2}}\gamma^5 
\partial_{\tilde\phi} \tilde\epsilon \\ \nonumber
\delta\tilde\Omega_G &=& -i \frac{1}{2}\partial_\mu G\gamma^\mu\tilde\epsilon
+ e^{-\frac{H+G}{2}} \partial_\phi\tilde\epsilon +i e^{ -\frac{H+G}{2}}  \gamma^5 
\partial_{\tilde\phi} \tilde\epsilon -\frac{1}{8} \gamma_{\rho\sigma}F^{\rho\sigma}
\tilde\epsilon.
\eeaj
We now can take the Kaluza-Klein ansatz for the spinor $\tilde\epsilon$
\bej{kaluza}
\tilde\epsilon(x, \phi,\tilde\phi) = \exp\left( -\frac{i}{2} ( \eta\phi + \tilde\eta\tilde\phi)\right)\tilde{\epsilon}(x).
\eej
Substituting this ansatz in the equations \eq{gravmod} we obtain
\beaj{gravmod1}
\delta\tilde\psi_\mu &=& \nabla_\mu \tilde\epsilon -\frac{i}{16} e^{-\frac{1}{2}(H-G)}\gamma_{\rho\sigma}
F^{\rho\sigma}\gamma_\mu \tilde\epsilon \\ \nonumber
i\delta\tilde\Omega_H &=& \partial_\mu H \gamma^\mu \tilde\epsilon
+ e^{-\frac{H+G}{2}} \eta \tilde\epsilon - ie^{-\frac{H-G}{2}}\gamma^5 
\tilde\eta \tilde\epsilon \\ \nonumber
i\delta\tilde\Omega_G &=& \partial_\mu G\gamma^\mu\tilde\epsilon
+ e^{-\frac{H+G}{2}} \eta\tilde\epsilon +i e^{ -\frac{H+G}{2}}  \gamma^5 
\tilde\eta \tilde\epsilon -\frac{i}{4} \gamma_{\rho\sigma}F^{\rho\sigma}
\tilde\epsilon.
\eeaj
These equations are now identical to the gravitino variation obtained by 
\cite{Liu:2004ru} for the case of 6d $(1,0)$ gravity. 
As a by product we have obtained the procedure to embed all the solutions
obtained by \cite{Liu:2004ru}  for 6d $(1,0)$ gravity in type IIB gravity. 
In appendix \ref{spinor}. we have explicitly written down 
two Killing spinors in $AdS_3\times S^4\times T^4$ in LLM coordinates
referred to by 
\bej{defkspiib}
\epsilon_1 = \frac{1}{2}( \tilde\epsilon + \tilde\epsilon^*), \qquad 
\epsilon_2 = \frac{i}{2}( \tilde\epsilon' + \tilde\epsilon^{\prime *}),
\eej
 where
 $\tilde\epsilon$ and $\tilde\epsilon'$ are given in \eq{full32} and \eq{full32a} respectively.

From the study of the supersymmetry algebra of type IIB gravity 
in \cite{Schwarz:1983qr} it is seen that  the action of 
anti-commutator of two supersymmetries contains a gauge transformation 
on the Neveu-Schwarz $B$-field. 
This implies that  under the presence of stretched 
strings the anti-commutator of the 
supercharges contains a term proportional to the winding charge of the 
stretched strings. We write this as
\bej{winchrg}
Q = \int d^9x \sqrt{-g} j^{M0}\omega_M,
\eej
where the winding current $j^{MN}$ is given by 
\bej{windcurr}
j^{MN}(x)  = \frac{R^2}{2\pi  \alpha'} \int_M d\tau d\sigma
( \partial_\tau X^M \partial_\sigma X^N - \partial_\tau X^N \partial_\sigma X^M )
\frac{\delta^{10} ( x - X(\sigma)) }{\sqrt{-g}}.
\eej
and $\omega_M$ is the resulting gauge transformation parameter from the 
anti-commutator of two supersymmetries.
Note that the winding  charge \eq{winchrg}
 is conserved only if $\omega_M$ is a closed 1-form
since $\nabla _M (j^{MN} \omega_M) = \nabla_M j^{MN}\omega_N + 
j^{MN}\nabla_M\omega_N =0$ only if $\nabla_M\omega_N - \nabla_N\omega_M$
vanishes. 
Thus what remains to be done is to determine the relevant gauge transformation
parameter which 
results from the action of the 
anti-commutator of two supersymmetries and show that that it is a closed form.
From the analysis done in appendix \ref{sugra}. and
appendix \ref{gaugetrans}. the relevant 
gauge transformation parameter is given by  
$\omega_\mu = i  (\bar\epsilon_1 \gamma_\mu \epsilon_2)$ where $\epsilon_1$ and
$\epsilon_2$ are given in \eq{defkspiib}.
In appendix \ref{gaugetrans}. we have explicitly evaluated $\omega_\mu$ and 
we see that it is a constant closed 1-from. The only non zero components of 
of $\omega_\mu$ is given by 
\beaj{GT}
\omega_{1}= \cos\chi, \qquad
\omega_{2}= \sin\chi,
\eeaj
where $\chi$ is the angle corresponding to the 
rotation degree of freedom in the $(1,2)$ plane.   
Since there is a freedom of  choosing the angle  $\chi$ 
the winding charge of a straight string
in the $x_1, x_2$ plane along
any direction is conserved and appears on the RHS of the supersymmetric
algebra.  All one has to do is to choose $\chi$ so that the 1-form $(\omega_1, \omega_2)$ is along the direction of the string. 
The magnitude of the winding charge along with its direction 
in the $(1,2)$ plane has $2$ independent components,
the straight stretched string  carries two additional central charges. 
For the giant magnon solution given in \eq{stline} the winding charge using
the definition in \eq{winchrg}, \eq{windcurr} is given by 
\bej{giwchrg}
\hat Q  =  \frac{R^2}{\pi\alpha'} \cos\theta_0 \hat x_2,
\eej
the $\hat x_2$ denotes the direction of the winding. 
Thus  subalgebra relevant for the giant magnons, 
$SU(1|1)\times SU(1|1)$ is therefore  extended by two additional central charges. 
From the discussion of the extended algebra in section 3, we see that the
extended charges also form a vector given by $ C_3 + i  C_4$. 
We now identify the central charges as 
\bej{strongident}
\frac{\hat Q}{2} =  ( C_3 + iC_4) .
\eej
The proportionality constant is fixed by the fact that the 
dispersion relation obtained from the BPS condition \eq{bpscond}
is consistent with the strong coupling dispersion relation obtained in \eq{dispersion}.
Since the magnons are 
straight stretched strings and carry the appropriate winding charges 
they are BPS. Therefore we can apply the BPS relation
\eq{bpscond} to derive the dispersion relation. For a single magnon
this gives
 \beaj{strongcdisp}
\Delta- J = ( C_1 + C_2) &=&
 \sqrt{ 1 + \left( \frac{R^2}{\alpha^{\prime } \pi}\right)^2 \cos^2 \theta_0 },
 \\ \nonumber
 &=&\sqrt{ 1 + \left( \frac{R^2}{\alpha^{\prime } \pi}\right)^2\sin^2 \frac{p}{2} }.
\eeaj
In the second line we have used the identification  given in \eq{basident} of
$\theta_0$ with the magnon momentum. 
Substituting the value of $R^2$ in terms of the D1, D5-brane charges we obtain
\bej{fulstrondisp}
\Delta - J = \sqrt{ 1 + \left(\frac{ g_6^2 Q_1Q_5} {\pi ^2}\right)  \sin^2 \frac{p}{2} }.
\eej
Thus  we see that at strong coupling
\bej{asymvalf}
f(\tilde\lambda) = \frac{g_6^2 Q_1Q_5}{\pi^2}, \qquad \tilde\lambda >>1.
\eej

\section{Discussion}

In this paper we have used the centrally extended $SU(1|1)\times SU(1|1)$ 
superalgebra with two more additional central charges to derive the dispersion relation
of magnons in the D1-D5 system. 
The derivation closely followed the derivation of the dispersion relation of
magnons in ${\cal N}=4$ Yang-Mills. 
This similarity suggests that just as ${\cal N}=4$ Yang-Mills is integrable at the 
planar limit, the 
D1-D5 system might be integrable for $Q_1Q_5>>1$. 
In fact the  classical string 
on $AdS_3\times S^3$ with Ramond-Ramond flux through
the $S^3$ is known  to have infinite set of non-local, commuting
conserved charges \cite{Bena:2003wd}. 
Thus using integrability along with the extended symmetry we have
found in this paper and proceeding algebraically 
along the lines of  \cite{Beisert:2005tm} it might be possible to obtain the 
S-matrix of this theory which will lead to the information about the  complete
spectrum in the large $J$ limit.  
Another approach is to look at the S-matrix of magnons in the 
strong coupling limit. 
As we have seen at strong coupling the giant magnon solution in the 
D1-D5 system is identical to that of ${\cal N}=4$ Yang-Mills.
Since the evaluation of the S-matrix for scattering of two magnons with momentum
$p_1$ and $p_2$ at strong coupling just depends on the 
classical solution, 
the S-matrix evaluated by \cite{Hofman:2006xt} applies for magnons in 
$AdS_3\times S^3$ as well. This is given by
\beaj{smatrix}
S (p_1, p_2) &=& \exp (i\delta) , \\ \nonumber
{\hbox{where}}\;\; \delta &=& -\frac{\sqrt{\lambda}}{\pi} \left(
\cos \frac{p_1}{2} -\cos \frac{p_2}{2} \right) 
\log \left( \frac{ \sin^2 \frac{p_1-p_2}{2} }{ \sin^2 \frac{p_1+p_2}{2} } \right) 
\eeaj
where ${\rm{sign}}( \sin\frac{p_1}{2})>0$ and ${\rm{sign}}( \sin\frac{p_1}{2}) >0$ and 
 $\lambda = g_6^2 Q_1Q_5$. 
It will be interesting to perform the sub-leading corrections to this 
S-matrix for the $AdS_3\times S^3$  case following \cite{Chen:2007vs}, 
since these depend on small fluctuations around the giant magnon background.
Here the fact that we are in the 
 $AdS_3\times S^3\times T^4$ background will play a role. The sub-leading
corrections to this S-matrix and the use of the extended symmetries we have
found in this work might help to determine the complete S-matrix.

Finally, we have
studied the extended supersymmetry of the giant magnons  following  the approach 
of \cite{Hofman:2006xt}. It will be interesting to study this issue using the 
more direct world sheet approach of \cite{Arutyunov:2006ak}.

\acknowledgments
The authors wish to thank Shiraz Minwalla and Yogesh Srivastava for discussionsand the HRI string group for useful comments.
We thank Yogesh Srivastava for a careful reading of the manuscript. 
 B. S  thanks the Centre for High Energy Physics, IISc, Bangalore
for hospitality during which the work was completed. 

\appendix

\section{Embedding  $(1,0)$ 6d supergravity in IIB supergravity} \label{sugra}

In this section we embed the solutions of $(1,0)$ 6d supergravity found in 
\cite{Liu:2004ru} in type IIB supergravity.
The strategy we follow is to rewrite the type IIB supersymmetry variations in 
terms of  supersymmetry variations of $(1,0)$ 6d supergravity given in \cite{Liu:2004ru}.
This allows us to easily solve for the Killing spinors for $AdS_3\times S^3$ in the 
LLM coordinates. 
We first choose the following convention of 10d gamma matrices.
\beaj{newgamma}
\Gamma^{\mu} &=& -i \gamma^{\mu}\otimes \varepsilon\otimes1\otimes\hat{\sigma_3}, 
\\ \nonumber
\Gamma^4 &=& i 1\otimes\varepsilon\otimes\tilde{\varepsilon}\otimes\hat{\varepsilon}, \\ \nonumber
\Gamma^5 &=& i 1\otimes 1\otimes \tilde{\sigma_1}\otimes\hat{\varepsilon}, \\ \nonumber
\Gamma^6 &=& i 1\otimes 1\otimes\tilde{\sigma_3}\otimes\hat{\varepsilon}, \\ \nonumber
\Gamma^7 &=& i 1\otimes \sigma_1\otimes\tilde{\varepsilon}\otimes 1, \\ \nonumber
\Gamma^8 &=& i 1\otimes \sigma_3\otimes \tilde{\varepsilon}\otimes 1, \\ \nonumber
\Gamma^9 &=& i 1\otimes \varepsilon\otimes1 \otimes\hat{\sigma_1}.
\eeaj
Here $\gamma^\mu$ with $\mu = 0, 1, 2, 3$ are $4\times 4$ gamma matrices in 
the Majorana representation, we write them down explictly below.
\beaj{def4dgamma}
\gamma^0 = -i \sigma_2\otimes 1, &\qquad&
\gamma^1 = \sigma^1\otimes 1, \\ \nonumber
\gamma^2 = \sigma^3\otimes\sigma^1, &\qquad&
\gamma^3 =  \sigma^3\otimes\sigma^3.
\eeaj
In \eq{newgamma} $\varepsilon$ refers to the following $2\times 2$ matrix
\bej{defepsil}
\varepsilon = \left(\begin{array}{cc}
0 & -1 \\ 1 &0 
\end{array}\right).
\eej
The $\tilde{}, \hat{}$ in \eq{newgamma} is used to keep track of which 2-component 
spinor the $2\times 2$ matrices act. 
Thus we have the following anti-commutation relations
\bej{antcomrel}
\{\Gamma^M, \Gamma^N\} = \eta^{MN}, \qquad
\{\gamma^\mu, \gamma^\nu \} = \eta^{\mu\nu}.
\eej
Note that all the 10 gamma matrices are purely imaginary. 
We now examine the supersymmetry variations \eq{a3} with this gamma matrix conventions.  Substituting the anstaz in \eq{a1} in the supersymmetry variations and 
setting the dilatino variation to zero gives
$F_{MNP}\Gamma^{MNP}\epsilon =0$, expanding this out we obtain
\beaj{dilexp}
F^{MNP}\Gamma_{MNP} &=&
\frac{3!}{2!} \left(
\Gamma_{\mu\nu} \Gamma_\phi F^{\mu\nu\phi} +
 \Gamma_{\mu\nu}\Gamma_{\tilde\phi}F^{\mu\nu\tilde\phi} \right),
 \\ \nonumber
 & =& 3 \left[-
\frac{1}{2}  e^{\frac{1}{2}( H+G)} \gamma_{\mu\nu} \Gamma_4 F^{\mu\nu}
e^{-(H+G)} 
-\frac{1}{2}e^{\frac{1}{2}(H-G)}\gamma_{\mu\nu} \Gamma_5 \tilde F^{\mu\nu} e^{-(H-G)}
\right], \\ \nonumber
&=& -\frac{3}{2} \left[
\gamma_{\mu\nu} \Gamma_4 F^{\mu\nu} - i \gamma^{\mu\nu}\gamma^5 F_{\mu\nu}
\Gamma_5 \right]e^{-\frac{1}{2}(H+G)} , \\ \nonumber
&=& -\frac{3}{2} \gamma_{\mu\nu}F^{\mu\nu} \Gamma_4 
\left( 1 + \Gamma^0\Gamma^1\Gamma^2\Gamma^3\Gamma^4\Gamma^5\right)e^{-\frac{1}{2}(H+G)}.
\eeaj
In the above manipulations we have used the fact that the $3$-form field strength
is self-dual in 6-dimensions  and 
\beaj{defgamm5a}
\gamma^5 = i \Gamma^0\Gamma^1\Gamma^2\Gamma^3, \qquad
\epsilon^{\mu\nu\rho\sigma}\gamma_{\rho\sigma} = 
-2i \gamma^{\mu\nu} \gamma^5.
\eeaj
Thus to set the dilatino variation to zero we need the following condition on the 
6d chirality condition on the spinor
\bej{6dchir}
\Gamma^0\Gamma^1\Gamma^2\Gamma^3\Gamma^4\Gamma^5\epsilon =
-\epsilon.
\eej
Since the spinor $\epsilon$ is a 10 Weyl spinor we also have the following condition
\bej{t4cond}
\Gamma^6\Gamma^7\Gamma^8\Gamma^9 \epsilon =-\epsilon.
\eej
We now look at the gravitino variation in the $4$ directions and obtain
\beaj{grav1}
\delta\psi_\mu &=&
\nabla_\mu\epsilon +\frac{1}{48}\frac{3}{2}
\gamma_{\rho\sigma}F^{\rho\sigma}e^{-\frac{1}{2}(H+G)} \Gamma_4 
\left( 1 + \Gamma^0\Gamma^1\Gamma^2\Gamma^3\Gamma^4\Gamma^5\right)
\Gamma_\mu\epsilon^*, \\ \nonumber
&=& 
\nabla_\mu\epsilon +\frac{1}{16} \gamma_{\rho\sigma}F^{\rho\sigma}e^{-\frac{1}{2}(H+G)} \Gamma_4\Gamma_\mu
\epsilon^* ,\\ \nonumber
&=& 
\nabla_\mu\epsilon -\frac{1}{16} \gamma_{\rho\sigma}F^{\rho\sigma}e^{-\frac{1}{2}(H+G)}
\gamma_\mu \tilde{\varepsilon}\hat{\sigma_1} \epsilon^*.
\eeaj
Note that the coefficients occurring in this equation are real due to our 
convention of the 10d gamma matrices.
Let us now look at the remaining components of the gravitino variation
\beaj{grav2}
\delta\psi_\phi &=&
\nabla_\phi\epsilon -\frac{1}{16} \gamma_{\rho\sigma} F^{\rho\sigma} 
\epsilon^* ,\\ \nonumber
&=& \partial_\phi \epsilon -\frac{1}{4}
\partial_\mu(H+G) e^{\frac{H+G}{2}} \Gamma_4\Gamma_\mu \epsilon
-\frac{1}{16} \gamma_{\rho\sigma} F^{\rho\sigma} \epsilon^*.
\eeaj
Similarly the last component of the gravitino variation becomes
\beaj{grav3}
\delta\psi_{\tilde\phi} &=&
\nabla_{\tilde\phi} \epsilon -\frac{1}{16} \gamma_{\rho\sigma} F^{\rho\sigma}
\Gamma_4\Gamma_5 e^{-G} \epsilon^*, \\ \nonumber 
&=& \partial_{\tilde\phi} \epsilon
-\frac{1}{4}\partial_\mu(H-G) e^{\frac{H-G}{2}}\Gamma_5\Gamma_\mu\epsilon
-\frac{1}{16} \gamma_{\rho\sigma} F^{\rho\sigma} e^{-G} \Gamma_4\Gamma_5
\epsilon^* .
\eeaj
Let us define the following linear combinations
\beaj{deflincomb}
\delta\Omega_H =
\delta\psi_\phi - e^G\Gamma_5\Gamma_4\delta\psi_{\tilde\phi}, \\ \nonumber
\delta\Omega_G = \delta\psi_{\phi} + e^G\Gamma_5\Gamma_4\delta
\psi_{\tilde\phi}.
\eeaj
we obtain the equations
\beaj{grav4}
\delta\Omega_H &=& -\frac{1}{2} \partial_\mu H\Gamma_4\Gamma^\mu\epsilon
+ e^{-\frac{H+G}{2}} \partial_\phi\epsilon - e^{-\frac{H-G}{2}} \Gamma_5\Gamma_4
\partial_{\tilde\phi} \epsilon, \\ \nonumber
\delta\Omega_G &=& -\frac{1}{2}\partial_\mu G \Gamma_4\Gamma^\mu 
+ e^{-\frac{H+G}{2}} \partial_\phi \epsilon + e^{-\frac{H-G}{2}} \Gamma_5\Gamma_4
\partial_{\tilde\phi} \epsilon  -\frac{1}{8} \gamma_{\rho\sigma} F^{\rho\sigma} 
\epsilon^* .
\eeaj
We now use the chirality condition  \eq{6dchir} and the representation of the
10d gamma matrices in \eq{newgamma} to write the above equations as
\beaj{grav5}
\delta\Omega_H &=& \frac{1}{2} \partial_\mu H\gamma^\mu \tilde{\varepsilon}\hat{\sigma_1}\epsilon + e^{-\frac{H+G}{2}} \partial_\phi\epsilon - ie^{-\frac{H-G}{2}} \gamma^5
\partial_{\tilde\phi} \epsilon, \\ \nonumber
\delta\Omega_G &=& \frac{1}{2} \partial_\mu G \gamma^\mu\tilde{\varepsilon}\hat{\sigma_1}\epsilon+ e^{-\frac{H+G}{2}} \partial_\phi \epsilon +  ie^{-\frac{H-G}{2}} \gamma^5
\partial_{\tilde\phi} \epsilon  -\frac{1}{8} \gamma_{\rho\sigma} F^{\rho\sigma} 
\epsilon^* .
\eeaj
Note that the coefficients of the above equations are also real, therefore we take
the real parts of the equations in \eq{grav2} and \eq{grav5}. This removes the 
complex conjugate operation on the spinor. We then consider the following spinor
\bej{ccspina}
\tilde\epsilon = \epsilon_R +  i (1\otimes\tilde{\varepsilon}\otimes\hat{\sigma_1}) \epsilon_R.
\eej
Here the subscripts stand for the real part of the spinor $\epsilon$, note that
the $1$ refers to the $8\times 8$ identity matrix.
The above form of $\tilde\epsilon$ is in fact  a reality condition on $\tilde\epsilon$, 
note that the reality condition does not involve the $4$-d spinor. Thus the 
$4$-d spinor is in general complex.  The reality condition explicitly 
is given by 
\bej{fulrealcond}
\tilde\epsilon + {\tilde\epsilon}^*  = i (1\otimes \tilde{\varepsilon}\otimes\hat{\sigma_1})(
\tilde\epsilon - {\tilde\epsilon}^* ).
\eej
The 6 dimensional chirality condition in \eq{6dchir} on $\tilde\epsilon$ reduces to the 
following 
\bej{6dchira}
\Gamma^0\Gamma^1\Gamma^2\Gamma^3\Gamma^4\Gamma^5 \tilde\epsilon 
=-\tilde\epsilon^*, 
\eej
while the chirality condition on the $T^4$ \eq{t4cond} directions remains 
\bej{t4chiral}
\Gamma^6\Gamma^7\Gamma^8\Gamma^9 \tilde\epsilon = -\tilde\epsilon.
\eej
We now write down the susy variation  equation for $\tilde\epsilon$  from
the equations \eq{grav2} and \eq{grav5}, we get  the following equations
\beaj{grav6}
\delta\tilde\psi_\mu &=& \nabla_\mu \tilde\epsilon +\frac{i}{16} \gamma_{\rho\sigma}
F^{\rho\sigma}\gamma_\mu \tilde\epsilon, \\ \nonumber
\delta\tilde\Omega_H &=& -i\frac{1}{2} \partial_\mu H \gamma^\mu \tilde\epsilon
+ e^{-\frac{H+G}{2}} \partial_\phi \tilde\epsilon - ie^{-\frac{H-G}{2}}\gamma^5 
\partial_{\tilde\phi} \tilde\epsilon, \\ \nonumber
\delta\tilde\Omega_G &=& -i \frac{1}{2}\partial_\mu G\gamma^\mu\tilde\epsilon
+ e^{-\frac{H+G}{2}} \partial_\phi\tilde\epsilon +i e^{ -\frac{H+G}{2}}  \gamma^5 
\partial_{\tilde\phi} \tilde\epsilon -\frac{1}{8} \gamma_{\rho\sigma}F^{\rho\sigma}
\tilde\epsilon.
\eeaj
We now can take the Kaluza-Klein anstaz for the spinor $\tilde\epsilon$
\bej{kkanstazs}
\tilde\epsilon(x, \phi,\tilde\phi) = \exp\left[ -\frac{i}{2} ( \eta\phi + \tilde\eta\tilde\phi)\right]
\tilde\epsilon(x)
\eej
Substituting this ansatz in the equations \eq{grav6} we obtain
\beaj{grav7}
\delta\tilde\psi_\mu &=& \nabla_\mu \tilde\epsilon +\frac{i}{16} \gamma_{\rho\sigma}
F^{\rho\sigma}\gamma_\mu \tilde\epsilon, \\ \nonumber
i\delta\tilde\Omega_H &=& \partial_\mu H \gamma^\mu \tilde\epsilon
+ e^{-\frac{H+G}{2}} \eta \tilde\epsilon - ie^{-\frac{H-G}{2}}\gamma^5 
\tilde\eta \tilde\epsilon, \\ \nonumber
i\delta\tilde\Omega_G &=& \partial_\mu G\gamma^\mu\tilde\epsilon
+ e^{-\frac{H+G}{2}} \eta\tilde\epsilon +i e^{ -\frac{H+G}{2}}  \gamma^5 
\tilde\eta \tilde\epsilon -\frac{i}{4} \gamma_{\rho\sigma}F^{\rho\sigma}
\tilde\epsilon.
\eeaj
The above supersymmetry variations 
 are identical to that obtained in \cite{Liu:2004ru} for the case
of $(1,0)$ 6d supergravity. Therefore we can use the methods discussed in 
\cite{Liu:2004ru} to find the Killing spinors. 
These equations determine the 4 dimensional  component of the 
10 dimensonal  Killing spinor. The remaining components are determined by the 
the conditions \eq{fulrealcond}, \eq{6dchira} and \eq{t4chiral}.

\subsection{Killing spinors in LLM coordinates} 
\label{spinor}

We will now find  the Killing spinors for  $AdS_3\times S^3$ in 
LLM coordinates. 
In these coordinates the solution is given by \cite{Liu:2004ru}
\beaj{ads3sol}
ds_6^2 &=& -h^2 ( dt + V_i dx^i)^2 
+ h^2( dy^2 + \delta_{ij} dx^idy^j) 
+ y ( e^G d\Omega_1^2 + e^{-G} d\tilde\Omega_1^2), \\
F_{(2)} &=& - 2 \left[
d\left( y e^G\right)\wedge ( dt + V)  + h^2 e^G *_3 d\left(
y e^{-G} \right) \right].
\eeaj
where
\bej{defhzva}
h^{-2} = 2y \cosh G, \qquad z \equiv \frac{1}{2} \tanh G, \qquad
dV = -\frac{1}{y} *_3 dz.
\eej
and $z$ satisfies the following equations
\beaj{zequa}
\left( \partial_i^2  + y\partial_y \frac{1}{y} \partial_y \right) z &=&0, \\ \nonumber
\partial_i z\partial_i z + \partial_y z\partial_y z &=& 
\frac{ ( 1-4z^2)^2}{4y^2}.
\eeaj
Let us first obtain the 4 dimensional part of the Killing spinor, we will then 
use the conditions \eq{fulrealcond}, \eq{6dchira} and \eq{t4chiral} to obtain
the full 10 dimensional Killing spinor. 
There are two choices for the Killing spinor of the above solution:
$\eta =1, \tilde\eta =-1$, or $\eta=-1, \tilde\eta =+1$ \cite{Liu:2004ru}. 
We now solve for the Killing spinor which has the condition
 $\eta=1, \tilde\eta =-1$. Substituting the solution given in \eq{ads3sol} into the 
the Killing spinor equation $\delta \tilde\Omega_H=0$  of
\eq{grav7} we obtain
\bej{simkileq}
\left( \sqrt{1+ e^{2G}} \gamma^3 + i \gamma^5 e^G + 1 \right) \tilde\epsilon =0.
\eej
Note that the above operator is a projector. To solve for the Killing spinor we
first choose the following ansatz
\bej{kilanstz}
\tilde\epsilon = \exp( i \delta \gamma^5\gamma^3) \tilde\epsilon_1, 
\qquad
\gamma^3 \tilde\epsilon_1 = -\tilde\epsilon_1.
\eej
Substituting the above ansatz in the Killing spinor equation given in 
\eq{simkileq} we obtain
\beaj{kilanstz1}
\left( \sqrt{1+ e^{2G}} \gamma^3 + i \gamma^5 e^G + 1 \right)
\left( \cosh\delta + i \sinh\delta \gamma^5\gamma^3\right) \tilde\epsilon_1 =0.
\\ \nonumber
\eeaj
Expanding the above equation  and equating the 
real and imaginary parts we obtain the following 
\bej{tan2}
\tanh\delta = \frac{\sqrt{1+ e^{2G}} -1}{e^G}, \qquad
\tanh\delta = \frac{e^G}{ \sqrt{1+ e^{2G}} +1}.
\eej
From these equations we obtain that
\bej{soldelta}
\sinh2\delta = \exp{G}.
\eej
We now fix the normalization of the spinor $\tilde\epsilon_1$:
Consider the following spinor bilinears
\bej{spinbil}
K^\mu = \bar{\tilde\epsilon}\gamma^\mu\tilde\epsilon, \qquad
L^\mu = \bar{\tilde\epsilon}\gamma^\mu \gamma^5 \tilde\epsilon.
\eej
Let $\tilde\epsilon_1 = \alpha \tilde\epsilon_0$
where $\tilde\epsilon_0^\dagger\tilde\epsilon_0 =1$. In the coordinate system
of the metric in \eq{ads3sol} we have 
$K^t =1, L_y = 1$, therefore we have the conditions
\beaj{normcond}
K^t= h\bar{\tilde{\epsilon}}\gamma^0 \tilde\epsilon = 1, \qquad
L_y = h\bar{\tilde{\epsilon}}\gamma^3\gamma^5 \tilde\epsilon =-1.
\eeaj
We  impose this normalization by first requiring the projector
\bej{exproj}
\gamma_0\gamma^3\gamma^5 \tilde\epsilon = \tilde\epsilon.
\eej
Now from the construction of $\tilde\epsilon$ in \eq{kilanstz} we see that
the above condition implies that 
\bej{proj1}
\gamma_0\gamma^5 \tilde\epsilon_1 =- \tilde\epsilon_1.
\eej
To fix the normalization constant $\alpha$ we evaluate the 
following scalar constructed out of spinor bilinears
\bej{scalbilin}
f_2= i \bar{\tilde{\epsilon}}\tilde\epsilon.
\eej
Substituting the form of the Killing spinor we obtain
\beaj{scalf2}
f_2 &=& i \tilde\epsilon_1\exp(-i\delta\gamma^3\gamma_5) \gamma^0 
\exp( i\gamma_5\gamma^3) \tilde\epsilon_1, \\ \nonumber
&=&  \alpha^2\sinh2\delta = \alpha^2 \exp(G).
\eeaj
From \cite{Liu:2004ru} we have 
$f_2 = \exp[(H+G)/2]$, therefore we obtain the value of the normalization constant
$\alpha$ as 
\bej{solalpha}
\alpha = \exp\left(\frac{H -G}{4}  \right).
\eej
Similar manipulations show that this normalization is consistent with
$K^t=1$ and $L_y=1$ in these cases one obtains the equation
\bej{consnorm}
h\cosh(2\delta) \alpha^2 =1.
\eej
The above equation can be easily shown to be true using the solution in 
\eq{ads3sol}. As part of the consistency requirement it can be shown using
simple gamma matrix manipulations that the 
other components of the vectors $K^\mu$ and $L_\mu$ vanish. 
Thus the Killing spinor is given by
\beaj{solkillspin}
\tilde\epsilon = \exp( \frac{H-G}{2} ) \exp( i \delta\gamma_5\gamma^3) \tilde\epsilon_0, 
\\ \nonumber
\gamma^3\tilde\epsilon_0 = -\tilde\epsilon_0, \qquad
\gamma_0\gamma^5\tilde\epsilon_0 =- \tilde\epsilon_0, 
\eeaj
or more explicitly the four component spinor is given by
\beaj{4compksp1}
\tilde\epsilon_4 &=&  e^{-\frac{i}{2} ( \phi -\tilde\phi) }e^{( \frac{H-G}{2} )}
( \cosh\delta - i \gamma^5\sinh\delta) \tilde\epsilon_0, \\ \nonumber
\tilde\epsilon_0 &=& \frac{1}{\sqrt{2}}
\left(
\begin{array}{c}
0\\1\\i\\0
\end{array}
\right).
\eeaj
It can be verified that the spinor in \eq{4compksp1} also satisfies the 
first equation in \eq{grav7}.
Note that there is a degree of freedom in choosing $\tilde\epsilon_0$ which corresponds 
to rotations in the $(1,2)$-plane. The constraints determining $\tilde\epsilon_0$
commutes with the rotation matrix $\gamma_1\gamma_2$, therfore we can 
also consider the following 
\bej{rotepnt}
\tilde\epsilon_0\rightarrow \exp({\chi\gamma_1\gamma_2}) \tilde\epsilon_0.
\eej
In \eq{4compksp1} 
we have reinstated the dependence on the coordinates $\phi, \tilde\phi$ and set the arbitrary constant phase  in $\tilde\epsilon_0$ to be zero. 
Performing the same analysis with $\eta =-1, \tilde\eta =1$ we obtain the 
following four component Killing spinor
\beaj{4compksp2}
\tilde\epsilon_4' &=& e^{\frac{i}{2} ( \phi -\tilde\phi) } e^{( \frac{H-G}{2} )}
( \cosh\delta - i \gamma^5\sinh\delta) \tilde\epsilon_0', \\ \nonumber
\tilde\epsilon_0' &=& \frac{1}{\sqrt{2}}
\left(
\begin{array}{c}
1\\0\\0\\i
\end{array}
\right).
\eeaj
Note that for this case compared to the case with $\eta =1, \tilde\eta =-1$ we have
$\delta \rightarrow -\delta$ and $\gamma^3\epsilon_0' =\epsilon_0'$ and 
$\gamma_0\gamma^5 \epsilon_0' = -\epsilon_0'$, since $L_y$ is now normalized to
$1$. 
There is again degree of freedom which corresponds to rotations in the $(1,2)$ plane
given by
\bej{rotation1}
\tilde\epsilon_0' \rightarrow  \exp({\chi\gamma_1\gamma_2}) \tilde\epsilon_0'.
\eej
With these ingredients  the full 10 dimensional 32 component
Killing  spinor which satisfies the conditions \eq{fulrealcond}, \eq{6dchira} and \eq{t4chiral} is given by
\beaj{full32}
\tilde\epsilon = \frac{1}{\sqrt{2}}\left(\begin{array}{c}
\tilde\epsilon_4' \\ \tilde\epsilon_4 \end{array}
\right)\otimes \left[ 
\left(\begin{array}{c}
1\\0 \end{array} \right)
\otimes
\left(\begin{array}{c}
0\\1 \end{array} \right)  + 
\left (\begin{array}{c}
0\\i \end{array} \right)
\otimes
\left(\begin{array}{c}
1\\0 \end{array} \right) \right].
\eeaj
The 6 dimensional chirality condition \eq{6dchira} is satisfied due to the 
property $\tilde\epsilon_4^* = i\gamma^5 \tilde\epsilon_4'$.  Furthermore the 
rotation degree of freedom
 in the $(1,2)$ plane
 \eq{rotepnt} and \eq{rotation1}
 has to be such that both $\tilde\epsilon_0$ and $\tilde\epsilon_0'$ has to be rotated
 by the same angle $\chi$.

Instead of working with the real part of the 10d spinor in \eq{grav5} we can also work
with the its imaginary part and construct the complex spinor
$\tilde\epsilon = \epsilon_I + i \tilde\varepsilon\hat\sigma_1 \epsilon_I$.
Then one obtains the same set of equations as given in \eq{grav7} with 
$F$ replaced by $-F$. Going through the same  procedure of solving for the 
complex spinor $\tilde\epsilon$ we obtain the following Killing spinor
\beaj{full32a}
\tilde\epsilon' = \frac{1}{\sqrt{2}}\left(\begin{array}{c}
\tilde\epsilon_4^{\prime *} \\ \tilde\epsilon_4^{ *} \end{array}
\right)\otimes \left[ 
\left(\begin{array}{c}
1\\0 \end{array} \right)
\otimes
\left(\begin{array}{c}
0\\1 \end{array} \right)  + 
\left (\begin{array}{c}
0\\i \end{array} \right)
\otimes
\left(\begin{array}{c}
1\\0 \end{array} \right) \right].
\eeaj
Note that the  4 component part of the full spinor for this case has
is complex conjugate compared the solution in \eq{full32}. 
This the because on replacing $F\rightarrow -F$ in \eq{grav7}, 
 $\epsilon_4^*$ and $\epsilon_4^{\prime *}$ are solutions.
The remaining components of the 32 dimensional spinor remains the same
since the construction of $\tilde\epsilon$ for this situation is same as the 
situation when one works only with the real part of $\epsilon$. 
It is easy to verify that the spinor in \eq{full32a} satisfies all the conditions
given in \eq{fulrealcond}, \eq{6dchira} and \eq{t4chiral}.
Thus we have constructed two Killing spinors \eq{full32} and \eq{full32a}
for $AdS_3\times S^3\times T^4$.

\section{Gauge transformation} \label{gaugetrans}

We now proceed to obtain the relevant gauge transformation parameter for the 
Neveu-Schwarz B-field appearing in the right hand side of the supersymmetry algebra. The relevant parameter is  given in  equation (2.34) of \cite{Schwarz:1983qr}
which is given by
\be \label{schwarz}
\Lambda_{\mu}^{\alpha}=A^{\alpha}_{\mu\rho}\xi^{\rho}-2i(V^{\alpha}_{+}\bar{\epsilon_{1}}\Gamma_{\mu}\epsilon_{2}^{*}+V^{\alpha}_{-}\bar{\epsilon_{1}}^{*}\Gamma_{\mu}\epsilon_{2}),
\ee
where $\epsilon_1, \epsilon_2$ are any two Killing spinors and $\Lambda_\mu^1=
\Lambda_\mu^{2*}$ and $A^1_{\mu\rho} = A^{2*}_{\mu\rho}$. 
The Neveu-Schwarz and the Ramond-Ramond  2-form are the real and imaginary
components of $A^\alpha_{\mu\rho}$. 
Here we have converted the gamma matrices used in \cite{Schwarz:1983qr} to our convention by setting $(\Gamma_{M})_{\rm{Schwarz}}=\Gamma_{M}\Gamma_{11}$.
$V^\alpha_{\pm}$ are related to the axion dilaton background. 
For constant dilaton backgrounds \footnote{The near horizon-geometry
of the D1-D5 system has a constant dilaton background.} 
one has 
\bej{defvval}
V^1_- = V^2_+,  \qquad \mbox{and} \;\; V^1_+ = V^{2*}_+.
\eej 
which are all constants. 
We now substitute the two Killing spinor solutions \eq{full32} and \eq{full32a}
into the expression for the gauge transformation \eq{schwarz}. 
For these solutions the 
32-component 10 dimension Killing spinor $\epsilon$ is related to 
$\tilde\epsilon$ by
\beaj{reltnont}
\epsilon_1 = \frac{1}{2} \left( \tilde\epsilon  + \tilde\epsilon^*\right), \qquad
\epsilon_2=  \frac{i}{2}\left( \tilde\epsilon' + \tilde\epsilon^{\prime *} \right).
\eeaj
Since  $\epsilon_1$ is purely real and $\epsilon_2$ is purely imaginary, the 
spinor bilinear $\bar\epsilon_1\Gamma_\mu \epsilon_2^*$ is purely
imaginary. From \eq{schwarz} and the conditions \eq{defvval} it is easy 
to see that the 
contribution to the real part of the gauge transformation parameter $\Lambda^\alpha_\mu$ from the spinor bilinear is proportional to 
$i \bar\epsilon_1\Gamma_\mu\epsilon_2$ while the contribution to the
imaginary part is given by $ \bar\epsilon_1\Gamma_\mu\epsilon_2$. 
Thus the gauge transformation parameter relevant for the
Neveu-Schwarz B-field is given by $i \bar\epsilon_1\Gamma_\mu\epsilon_2$. 

We now evaluate the spinor bilinear relevant for the gauge transformation
parameter and show that it is a constant. 
We have
\bej{gtparf}
i(\bar\epsilon_{1}\Gamma_{\mu}\epsilon_{2}) =
\rm{Re}\left( \tilde\epsilon_4^T\gamma_0\gamma_\mu \tilde \epsilon_4 
+ \tilde\epsilon_4^{\prime T}\gamma_0\gamma_\mu \tilde\epsilon_4' \right).
\eej
Here we have used \eq{reltnont} and substituted the solutions for
$\epsilon$ and $\epsilon'$ given in \eq{full32} and \eq{full32a}. 
Let us first evaluate 
$\rm{Re}( \tilde\epsilon_4^T\gamma_0\gamma_\mu \tilde \epsilon_4)$
at say $\phi=0, \tilde\phi =0$ \footnote{The giant magnon is located at a 
definite point along these directions. }.  
Using the   the expression given in \eq{4compksp1} we find:
\ben \label{gauge1}
{\tilde{\epsilon}}^T_4\gamma_0\gamma_{0}\tilde{\epsilon}_4&=&
\tilde{\epsilon}_4^T\gamma_0\gamma_{3}\tilde{\epsilon}_4=0, \nonumber \\
\tilde{\epsilon}_4^T\gamma_0\gamma_{i}\tilde{\epsilon}_4&=& e_{i}^{\hat{i}}\tilde{\epsilon}_4^T
\gamma_0\gamma_{\hat{i}}\tilde{\epsilon}_4, \nonumber \\
&=& e_{i}^{\hat{i}}\exp{(\frac{H-G}{2})}\tilde{\epsilon}_{0}^T\gamma_0\gamma_{\hat{i}}(\cosh2\delta+i\sinh2\delta\gamma^{5}\gamma^{3})\tilde{\epsilon_{0}}.
\een
But using the solution in  \eq{4compksp1} one can show that $\tilde{\epsilon}_{0}^T\gamma_0\gamma_{\hat{i}}\gamma^{5}\gamma^{3}\tilde{\epsilon_{0}}=0$. We also have 
\ben \label{gauge2}
\tilde{\epsilon}_{0}^T\gamma_0\gamma_{\hat{1}}\tilde{\epsilon_{0}}=1, \qquad
\tilde{\epsilon}_{0}^T\gamma_0\gamma_{\hat{2}}\tilde{\epsilon_{0}}=i.
\een
 Therefore 
\ben \label{gauge3}
\rm{Re}(\tilde{\epsilon}_4^T\gamma_0\gamma_{\hat{1}}\tilde{\epsilon}_4)&=&\exp(\frac{H-G}{2})\cosh2\delta=h^{-1}, \nonumber \\
\rm{Re}(\tilde{\epsilon}_4^T\gamma_0\gamma_{\hat{2}}\tilde{\epsilon}_4)&=&0.
\een
We then have
\ben \label{gauge4}
\rm{Re}(\tilde{\epsilon}_4^T\gamma_0\gamma_{1}\tilde{\epsilon}_4)&=&
e_{1}^{\hat{1}}\rm{Re}(\tilde{\epsilon}_4^T\gamma_0\gamma_{\hat{1}}\tilde{\epsilon}_4), \nonumber \\
&=&h \times h^{-1}=1. \nonumber \\
\rm{Re}(\tilde{\epsilon}_4^T\gamma_0\gamma_{2}\tilde{\epsilon}_4)&=&
e_{2}^{\hat{2}}\rm{Re}(\tilde{\epsilon}_4^T\gamma_0\gamma_0\gamma_{\hat{2}}\tilde{\epsilon}_4), \nonumber \\
&=&0.
\een
A similar calculation yields
\beaj{gauge5}
\rm{Re}(\tilde{\epsilon}_4^{\prime T}\gamma_0\gamma_{1}\tilde{\epsilon}_4') =1, \qquad
\rm{Re}(\tilde{\epsilon}_4^{\prime T}\gamma_0\gamma_{2}\tilde{\epsilon}_4')=0.
\eeaj
 All the remaining components of the above bilinear vanish. 
Combining \eq{gauge4} and \eq{gauge5} the relevant gauge parameter is given by
\beaj{finalgauge}
- i \rm{Re}( \epsilon_1^T\gamma_0\gamma_1\epsilon_2 ) =  1, \qquad
-i  \rm{Re}( \epsilon_1^T\gamma_0\gamma_2\epsilon_2 ) =0.
\eeaj
But since we have the freedom of rotation of the  solution in the $(1,2)$ given by
\eq{rotepnt} and \eq{rotation1} we can rotate the above gauge parameter to 
point along any direction in the $(1,2)$ plane. We refer to the gauge 
parameter as $\omega_\mu$ and the non-vanishing components 
are given by
\bej{fintgauge}
\omega_1 = \cos\chi, \qquad \omega_2 =\sin\chi.
\eej
where $\chi$ is a constant. 
 \bibliography{magnon}
\bibliographystyle{JHEP}

\end{document}